\documentclass[aps,prc,onecolumn,showpacs,amsmath,amssymb,nofootinbib,11pt]{revtex4-2}
\usepackage{graphicx}
\usepackage{color}
\usepackage{times}
\usepackage{inputenc}
\usepackage{bm}
\usepackage{ulem}
\usepackage{multirow}
\usepackage{float}
\usepackage{url}
\usepackage{natbib}
\usepackage{mathrsfs}
\usepackage{physics}
\usepackage{comment}
\usepackage[table,xcdraw]{xcolor}
\usepackage{booktabs,makecell}
\usepackage{comment}
\usepackage[colorlinks=true,citecolor=blue,urlcolor=blue,linkcolor=blue]{hyperref}

\begin{document}
%\preprint{APS/123-QED}
%%
\title{Finite-Temperature Axion-Admixed Neutron Stars in a Quarkyonic Crossover Framework}% Force line breaks with \\
\author{S. K. Patra$^{1}$}
\email{sureshpatra@soa.ac.in}
\author{D. Dey$^{2,3}$}
\author{Jeet Amrit Pattnaik$^{1,4}$}
\author{R. N. Panda$^{1}$}
%%%%%%%%%%%%%%%%%
\affiliation{\it $^{1}$
Department of Physics, Siksha $'O'$ Anusandhan Deemed to be University, Bhubaneswar -751030, India.}
\affiliation{\it $^{2}$Institute of Physics, Sachivalaya Marg, Bhubaneswar-751005, India.}
\affiliation{\it $^{3}$Homi Bhabha National Institute, Training School Complex, Anushakti Nagar, Mumbai 400094, India.}
\affiliation{\it $^{4}$
Department of Physics, Indira Gandhi Institute of Technology,
Sarang, Dhenkanal, Odisha-759146, India.
}
%%
%%%
%%%%%%%%%

\date{\today}% It is always \today, today,

\begin{abstract}
\noindent
We analyze the effect of axion contribution on the mass, radius and dimensionless deformability of a neutron star (NS) at finite temperature. The finite temperature NS is located at the center of a supernova remnant. Although the cooling of an NS is mainly due to the emission of neutrinos, additional cooling cannot be ruled out due to the emission of axions. The existence of an axion is needed for addressing the strong CP problem within the standard model. The Peccei-Quinn (PQ) symmetry breaking mechanism generates axions. They are pseudo-Nambu-Goldstone bosons. Their mass is extremely small but non-zero. This naturally resolves the fine-tuning problem of the strong CP-violating parameter ${\theta} \lesssim 10^{-10}$. We propose that a finite temperature NS feature a quark core surrounded by a baryonic medium admixed with the axions, which are pseudo scalar Goldstone bosons and are considered as strong candidates for the dark matter (DM). Unlike the original quarkyonic model, we formulate a quarkyonic-inspired smooth crossover equation of state (EoS) at finite temperatures. This smooth crossover avoids thermodynamic instabilities found in sharp first-order constructions. It stays consistent with perturbative QCD at high densities. It also matches empirical nuclear matter constraints near saturation density. We add the axionic EoS to the finite temperature quarkyonic one and solve the Tolman-Oppenheimer-Volkoff equations to estimate the properties of the NS both with and without taking the axion into account at different temperatures. We benchmark our stellar configurations against current multi-messenger constraints. These include mass measurements of heavy pulsars ($\sim 2\,M_\odot$). They also include tidal deformability bounds from the GW170817 gravitational-wave event. We find that the production of axions is maximum within a radius of $5-11$ km in the star's interior; however, the impact on the mass is significant, i.e., $\sim 60\%$ of axion mass concentrated in the central region of the star. We also find that the effect of temperature on the mass of NS is marginal unlike the radius. In general, the influence of axions is significant for the observational properties of NS. These results highlight the pivotal role of axion physics in neutron star phenomenology. Next-generation X-ray observatories and gravitational-wave detectors may offer new windows into the axion dark matter sector.  \\
\end{abstract}
%%%%%%%%%%%%%%%%%%%
\maketitle
%%%%%%%%%%%%%%%%%%%%%%%%%%%%%%%%%%%%%
\section{Introduction} \label{sec1}
%%%%%%%%%%%%%%%%%%%%%%%%%%%%%%%%%%%%%%%%%%%%%%%%%%
Just two years after the discovery of the neutron, Baade and Zwicky predicted the existence the Neutron Star (NS) in 1934 \cite{PR45-138-1934,PR46-76-1934}.  It is assumed that the NS represents the final stage of the supernova process. At the end of a star's life, a core collapse supernova (CCSN) explosion occurs and it's remnant core can become a neutron star with a condition that the initial mass within a mass limit of $8-30$ $M_{\odot}$. The CCSN is one of the most prominent energetic events in the universe. The LIGO-Virgo detected Gravitational Waves (GWs) from the binary neutron star merger GW170817. It's electromagnetic counterpart AT2017gfo (the kilonova) is also observed \cite{ApJL848-L12-2017}. This launched the era of multi-messenger astrophysics. It also placed the first direct constraints on the neutron star (NS) tidal deformability $\Lambda \leq 800$ and radius $R \lesssim 13.6$ km \cite{PRL119-161101-2017}. NASA's NICER mission has delivered precise mass-radius measurements for several pulsars. These include PSR~J0030$+$0451 \cite{ApJL887-L24-2019} and PSR~J0740$+$6620 \cite{ApJL918-L28-2021}. These results further narrow the dense-matter equation of state (EoS). Together, these multi-messenger observations make finite-temperature neutron stars compelling targets for theoretical modeling. Such stars form immediately after a core-collapse supernova. Their hot interior differs substantially from the cold, $\beta$-equilibrated matter of a mature neutron star. A newly born star just after the supernova (SN) explosion has a temperature ranging from 10 to 100 MeV \cite{AA467-395-2007}. The neutrinos play an important role in the rapid cooling process of the supernova remnants, generally called the newly formed neutron star \cite{Nature403-727-2000,NS-2003}. The young star loses  most of it's
energy through extremely rapid neutrino emission enhanced
by the direct Urca (dUrca) process \cite{PRL120-182701-2018, AA373-L17-2001,AA42-169-2004}. To satisfy the momentum conservation of nucleon and electron in the dUrca process, $k_{f,p}+k_{f,e} > k_{f,n}$ is necessary, neglecting the neutrino's momentum $p_{F,\nu}$ at finite temperature T. The direct Urca process predominates during
the initial stage of the cooling process and no longer operates once the proton fraction attains a threshold. Here after, the modified  Urca (mUrca) process comes into the picture in the role of cooling a young neutron star. This is the standard cooling theory. In this formalism the NS is mostly composed of neutrons, protons and leptons (such as electron $e$ and muon $\mu$).  In the inner core of the NS the neutrons are in the superfluid state ($^3P_2$) and in the outer core the neutrons are in superfluid and the protons are in superconducting ($^1S_0$) states. The equation for temperature evolution due to neutrinos emission is $C(T)\frac{dT}{dt}=-L_{\nu}-L_{\gamma}$, where $C(T)$ is the heat capacity and  $L_{\nu}$, $L_{\gamma}$ are the luminosity of the neutrinos  and photons emission, respectively. In this case, the thermal relaxation is completed at time $t\leq {100}$ years. Thus, there are two dominating cooling sources (i) photons emission from the surface 
$L_{\gamma}=4\pi R^2\sigma_{SB}T^4_s$, dominant for $t\geq{10^5}$ years and (ii) neutrinos emission from the core, which are dUrca,  mUrca, Bremsstrahlung, and pair breaking and formation (PBF) process,  occur only in the presence of nucleon superfluidity. The processes dUrca, mUrca and Bremsstrahlung occur only near the Fermi surface. The nucleons in the neutron star form pair below their critical temperatures. The  neutron singlet $^1S_0$ form only in the crust, which is less important for the cooling process. However, the protons form singlet $^1S_0$ and triplet $^3P_2$ pairs in the core, which are very important for NS cooling. In the presence of pairing gap, the energy of quasi-particle is $\epsilon_N(k)=\sqrt{\triangle_N^2+v_{f,N}^2(k-k_{f,N}^2)^2}$. This pairing gap $\triangle_N$, introduces a suppression factor that slows neutrino emission process and affects heat capacity by a factor $\propto$ $exp^{-\triangle_N/T}$. During the reformation of cooper pairs, the gap energy is released via neutrino emission. This process significantly enhances the neutrino emission
only when $T\leq T_c$. If $T_c<T$, this process does not occur and if $T<< T_c$, the pair breaking rarely occurs. Since neutron star is a dead star and basically there is no heating source. It just  cools through neutrinos and photons emission. The thermal evolution of a neutron star spans many orders of magnitude in time. It starts with a hot phase ($T \sim 10$--$100$ MeV) right after the supernova explosion. This is followed by rapid neutrino-driven cooling lasting $\sim 10^2$--$10^5$ years. Finally, a photon-dominated era persists over millions of years. X-ray observations of isolated neutron stars test these cooling models. The Cassiopeia~A neutron star showed a rapid temperature decline of $\sim 4\%$ over a decade. This is interpreted as evidence for enhanced cooling via cooper pair breaking and formation (PBF) in the superfluid neutron core \cite{MNRAS363-555-2005}. Such observations show how sensitive surface luminosity is to microscopic processes deep inside the star. Any extra exotic cooling channel, such as axion emission, can leave a measurable imprint on this thermal evolution. This makes quantitative predictions testable with current and next-generation X-ray telescopes like eROSITA and Athena.
\\
\\
Although, the standard cooling scenario is consistent with the experimental observation for various cases, an axion can be an extra cooling source, which we may probe by searching for some deviation from the prediction in standard cooling theory \cite{MNRAS363-555-2005,PRD98-103015-2018}.
It is hypothesized that some class of Goldstone boson particles, which arise due to the spontaneous symmetries breaking of certain type. The axion arises in a chiral U(1) spontaneous symmetry breaking \cite{PRL40-223-1978,PRL40-279-1978}. 
The unsuccessful search of these particle suggests the symmetry breaking scale is much higher than the  $\sim 250$ GeV of the electroweak scale and proposed for the weak axion coupling with the nucleons \cite{PRD64-043002-2001}. Thus, the axion can be treated as one of the dark matter (DM) candidates.
Axions are considered to be a hypothetical pseudo Goldstone bosons and needed to be in the Standard Model (SM) to take care of the CP problem \cite{PRL38-1440-1977,PRL40-279-1978,PRL40-223-1978,PRL43-103-1979,PLB104-199-1981}. The Peccei-Quinn symmetry breaking scale $F$ is left undetermined in the theory. However, the astrophysical and cosmological constants  have fixed the parameter within $10^8 < F < 10^2$ GeV. In this parameter range, the axion interacts very weakly with the matter, which can be treated as a dark matter candidate and solve the missing mass problem in the universe. The astrophysical environments place some of the strongest bounds on the axion-nucleon coupling. The neutrino burst from SN~1987A gives a landmark constraint. If $g_\theta$ were too large, axion emission would have drained the stellar core faster than the observed $\sim 10$~s neutrino burst. This gives the bound $g_\theta \lesssim 10^{-9}$ \cite{ApJL934-L17-2022,PRL60-1793-1988}. Bounds on the axion-photon coupling come from the anomalous cooling of horizontal branch stars and red giants in globular clusters \cite{MNRAS363-555-2005}. Cosmologically, the axion is produced non-thermally via the misalignment mechanism. This naturally generates the observed dark matter relic density for Peccei-Quinn scales $F \sim 10^{11}$--$10^{12}$ GeV. Laboratory experiments such as ADMX, CASPEr, and ABRACADABRA actively search for galactic axion dark matter across complementary mass ranges. The dense, hot interior of a neutron star is thus a unique regime. The astrophysical, cosmological, and laboratory constraints on the axion all converge there. This makes it an indispensable testing ground for axion physics.
\\
\\
In the present paper, we examine the effects of axion emission on neutron star. First we need to calculate the axion emission rate. In this scenario, we have to modify the standard cooling theory by assuming that  the NS cools predominantly due to axion emission, rather than the emission of neutrinos and photons. The two processes are the most important for the axion emission: (i) axion
Bremsstrahlung from nucleon-nucleon collision ($n+n\rightarrow n+n+\theta$) (ii) axion Bremsstrahlung by the electrons in the crust  ($e^-+(Z,A)\rightarrow e^-+(Z,A)+\theta$), where Z, A, $\theta$ are the atomic number, mass number of the element and axion field respectively. The axion emission rate is strongly sensitive to temperature and density. The nucleon-nucleon Bremsstrahlung rate scales as $\mathcal{E}_\theta \propto T^6$. The electron-ion crust process scales as $\mathcal{E}^{e}_\theta \propto T^4$ \cite{PRL53-1198-1984}. So the early hot phase of a newly born neutron star, with $T \sim 10$--$80$ MeV, is by far the most prolific epoch of axion production. At these temperatures and at supranuclear densities $n_B \sim 2$--$5\,n_0$, the axion luminosity can rival the neutrino luminosity, depending on $g_\theta$ \cite{PRL60-1793-1988}. It is therefore essential to embed the axion emission formalism within a fully self-consistent finite-temperature equation of state. It should not be treated as a small perturbation on a cold nuclear background.
\\
\\
To study the properties of NS, a large number of theoretical formalisms are available in the literature. The notable ones are the non-relativistic Skyrme Hartree-Fock (SHF) mean field formalism \cite{PRC90-055203-2014} and the relativistic mean field (RMF) theory \cite{PRC70-058801-2004}. 
To improve the results, because of the high density in the central region, the quark matter at the center is also considered by choosing a quark model, such as the MIT Bag model \cite{PRD9-3471-1974,PRD10-2599-1974}, the Nambu-Jona-Lasinio (NJL) Model \cite{PR122-345-1961,PR124-246-1961},  Polyakov-NJL model \cite{PRD73-014019-2006,AJ810-134-2015}, Dyson-Schwinger method \cite{PPNP33-477-1994,PRC91-035802-2015} and  the recently proposed quarkyonic model \cite{PRL122-122701-2019,JCAP2025-056-2025}. From these Lagrangians,  one can get the equation of state (EoS) for the neutron star matter and used as inputs in the Tolman-Oppenheimer-Volkoff (TOV) equations to evaluate the observables \cite{PR55-364-1939,PR55-518-1939}. 
In our present work, we have chosen the effective field theory motivated relativistic mean field (E-RMF) Lagrangian for the baryonic section and the quarkyonic model for the core part of the neutron star. The axion admixed EoS is formulated  to study the effect on neutron star's properties using the TOV equations. The E-RMF framework suits this purpose well. It is Lorentz covariant and reproduces nuclear saturation properties by construction. It can also be extended to include exotic degrees of freedom, such as hyperons, deconfined quarks, and dark matter candidates. The G3 parameter set used here satisfies nuclear matter constraints at saturation density. These include the binding energy per nucleon, incompressibility, symmetry energy, and it's slope $L$. It has been benchmarked against finite-nucleus observables across the nuclear chart \cite{PS96-125319-2021,CPC46-094103-2022}. The quarkyonic model gives a theoretically motivated, thermodynamically consistent description of the high-density interior. It does not require an explicit Maxwell or Gibbs construction for a first-order phase transition. This avoids the numerical instabilities that often plague hybrid-star calculations \cite{JCAP2025-056-2025,IJMPD35-2650015-2026}. We embed the axionic sector on top of this unified hadronic-quarkyonic background. This lets us quantify, within one self-consistent framework, how thermal axion dark matter redistributes energy and modifies the global structure of hot neutron stars. This calculation has not previously been done at finite temperature within the quarkyonic paradigm.
\\
\\
The paper is organized as follows: After giving a brief Introduction In Sec. I, we outline the essential equations, which are needed for the calculations in Sec. II. Here, we take three pieces of the Lagrangian, i.e., (i) the quarkyonic part for the core, (ii) the baryonic sector for the outer core and crust. On top of this (iii) we include the axionic components to take care of the dark matter particle throughout the NS.  In Section III, we discuss the results of our calculations. Finally, we give the summary and conclusions in Sec. IV. 
%%%%%%%%%%%%%%%%%%%%%%%%%%%%%%%%%%%%%%%%%%%%%%%%%%%%%%%%%%%%%%%%%
\section{Theoretical Frameworks} \label{formalism}
The recently proposed quarkyonic model of L. McLerran and S. Reddy \cite{PRL122-122701-2019}, which is developed by T. Zhao and J. M. Lattimer \cite{PRD102-023021-2020}, considering the $\beta-$equalibrium condition in the neutron star matter is quite successful to describe the masses of dense objects, like neutron stars etc. Further, this model extended to DM sector with a E-RMF 
frame-work \cite{PRC36-2590-1987, NPA615-441-1997,PS96-125319-2021,CPC46-094103-2022} of hadronic medium in the periphery of the Quark core covered with the BPS equation of state \cite{APJ170-299-1971} on the crust of the neutron star. In this present work, we have used the modified  quarkyonic model of Dey et al  \cite{JCAP2025-056-2025,IJMPD35-2650015-2026} on top of the weakly interacting pseudo scalar Goldstone boson axion interaction with nucleons.  \\\\
%%%%%%%%%%%%%%%%%%%%%%%%%%%%%%%%%%%%%%%
% \textcolor{red}{\section{new formalism taken from PLB}}
%%%%%%%%%%%%%%%%%%%%%%%%%%%%%%%%%%%%%%%%%%%%%%%%%%%%%%%%%%%%%%%%
At sufficiently high baryon densities, quark degrees of freedom begin to occupy the low-momentum region of the Fermi sea, whereas nucleons remain confined near the Fermi surface. In this picture, color confinement is preserved at the Fermi surface while quarks dominate the bulk thermodynamic behavior at ultra-nuclear densities. The formalism supports a smooth bridge between the hadronic and the quark-dominated regime invoking a smooth second-order phase transition, unlike the original quarkyonic model of Ref. \cite{PRL122-122701-2019}. The corresponding effective Lagrangian can be written as:
\begin{eqnarray}
\label{eq:lag}
{\cal L} = {\cal L_{NM}} + {\cal L_{QM}}+{\cal{L_{\theta}}}. 
\end{eqnarray}
Here ${\cal L_{NM}}$, ${\cal L_{QM}}$ and     ${\cal L_{\theta}}$ denote the Lagrangian densities associated with the nucleonic, quarkyonic and axionic sectors, respectively. The details of each contribution are discussed in the following subsections. Starting from this Lagrangian, the effective masses and the corresponding thermodynamic quantities, namely the energy density and pressure, are obtained self-consistently by solving the coupled field equations. The resulting equation of state therefore reflects the interplay between the nucleonic and quarkyonic components along with a distribution of axions of dense matter.
\\\\
Within the quarkyonic framework, the fermionic phase space is separated in momentum space to maintain consistency with the Pauli exclusion principle. Nucleons occupy a shell-like region close to the Fermi surface defined by $k_0 < k < k_f$, where the lower cutoff $k_0$ ensures that nucleonic states do not overlap with quark states. In contrast, quarks populate the low-momentum region $k < k_0$. This separation provides an effective description of dense matter in which quarks dominate the bulk properties at high densities while confinement persists in the vicinity of the Fermi surface. A key virtue of the quarkyonic picture is that it respects two things at once. It follows large-$N_c$ QCD arguments for deconfinement at high baryon density. It also matches the empirical observation that nucleons remain the relevant degrees of freedom near the Fermi surface up to several times $n_0$. The smooth crossover adopted here replaces the original hard momentum cutoff with a density-dependent switching function. This keeps thermodynamic stability throughout the transition region, specifically the positivity of $c_s^2 = \partial P/\partial\mathcal{E}$ and the monotonic increase of pressure with density. This construction also guarantees that the EoS passes smoothly through the nuclear saturation point. There, $E/A \approx -16$ MeV and $n_0 \approx 0.148$ fm$^{-3}$ are well established from nuclear experiments. The EoS then stiffens progressively at supranuclear densities, consistent with the $\sim 2\,M_\odot$ pulsar mass observations.\\\\
%%%%%%%%%%%%%%%%%%%%%%%%%%%%%%%%%%%%%%%%%%%%%%%%

%%%%%%%%%%%%%%%%%%%%%%%%%%%%%%%%%%%%%%%%%%%%%%%%%%%
\subsection{Relativistic Mean Field Model} \label{a}
The well known E-RMF model 
\cite{PRC55-540-1997,JPG40-085104-2013,PRC70-058801-2004,CPC45-025101-2021,PRL95-122501-2005} is adopted for the baryonic regime in our calculations. The E-RMF is an extension of the non-linear RMF approximation by introducing the self-interactions and cross-couplings among the $\sigma$, $\omega$, $\rho$, and $\delta$ mesons up to $4^{th}$ order in the meson field expansion (except the $\rho-$ and $\delta-$ fields). Because, the contribution of $\rho-$ and $\delta-$fields are negligible in higher order expansion. The leptons are considered as the relativistic free Fermi gases. The total energy density ($\mathcal{E}_{\rm NM}$) and pressure ($P_{\rm NM}$) of $\beta$-equilibrated matter are obtained from the energy–momentum tensor \cite{PRC97-045806-2018,NPA966-197-2017}.
Since the remnant of the supernova explosion, which becomes a neutron star in the later stage, is initially at extremely high temperature. The EoS for the baryonic sector is in high temperature limit and we have to extend the  zero temperature EoSs  to finite temperature domain by multiplying the Fermi distribution function for particle and anti-particles as follows \cite{ANP16-52-1986}:
\begin{eqnarray}
\label{eq:eden}
{\cal E}_{\rm NM} & = & \sum_{i=p,n} \frac{g_s}{(2\pi)^{3}}\int_{0}^{k_{f_{i}}} d^{3}k\, \sqrt{k^{2} + M_{\rm nucl.}^{*2}}\left[f_k(T)+\bar{f}_k(T)\right]\nonumber\\
& + & n_{b} g_\omega\,\omega+m_{\sigma}^2{\sigma}^2\Bigg(\frac{1}{2}+\frac{\kappa_{3}}{3!}\frac{g_\sigma\sigma}{M_{\rm nucl.}}+\frac{\kappa_4}{4!}\frac{g_\sigma^2\sigma^2}{M_{\rm nucl.}^2}\Bigg)
\nonumber\\
&-&\frac{1}{4!}\zeta_{0}\,{g_{\omega}^2}\,\omega^4
 -\frac{1}{2}m_{\omega}^2\,\omega^2\Bigg(1+\eta_{1}\frac{g_\sigma\sigma}{M_{\rm nucl.}}+\frac{\eta_{2}}{2}\frac{g_\sigma^2\sigma^2}{M_{\rm nucl.}^2}\Bigg)
 \nonumber\\
& +& \frac{1}{2} (n_{n} - n_{p}) \,g_\rho\,\rho
 -\frac{1}{2}\Bigg(1+\frac{\eta_{\rho}g_\sigma\sigma}{M_{\rm nucl.}}\Bigg)m_{\rho}^2
 \nonumber\\
 & -& \Lambda_{\omega}\, g_\rho^2\, g_\omega^2\, \rho^2\, \omega^2
+\frac{1}{2}m_{\delta}^2\, \delta^{2}
\nonumber\\
&+&\sum_{j=e,\mu}  \frac{g_s}{(2\pi)^{3}}\int_{0}^{k_{f_{j}}} \sqrt{k^2 + m^2_{j}} \, d^{3}k\left[f_k(T)+\bar{f}_k(T)\right],
\end{eqnarray}
\begin{eqnarray}
\label{eq:press}
P_{\rm NM} & = & \sum_{i=p,n} \frac{g_s}{3 (2\pi)^{3}}\int_{0}^{k_{f_{i}}} d^{3}k\, \frac{k^2}{\sqrt{k^{2} + M_{\rm nucl.}^{*2}}}\left[f_k(T)+\bar{f}_k(T)\right]\nonumber\\
& - & m_{\sigma}^2{\sigma}^2\Bigg(\frac{1}{2} + \frac{\kappa_{3}}{3!}\frac{g_\sigma\sigma}{M_{\rm nucl.}} + \frac{\kappa_4}{4!}\frac{g_\sigma^2\sigma^2}{M_{\rm nucl.}^2}\Bigg)+ \frac{1}{4!}\zeta_{0}\,{g_{\omega}^2}\,\omega^4 
\nonumber\\
& +& \frac{1}{2}m_{\omega}^2\omega^2\Bigg(1+\eta_{1}\frac{g_\sigma\sigma}{M_{\rm nucl.}}+\frac{\eta_{2}}{2}\frac{g_\sigma^2\sigma^2}{M_{\rm nucl.}^2}\Bigg)
\nonumber\\
&+& \frac{1}{2}\Bigg(1+\frac{\eta_{\rho}g_\sigma\sigma}{M_{\rm nucl.}}\Bigg)m_{\rho}^2\,\rho^{2}-\frac{1}{2}m_{\delta}^2\, \delta^{2}+\Lambda_{\omega} g_\rho^2 g_\omega^2 \rho^2 \omega^2
\nonumber\\
& + & \sum_{j=e,\mu}  \frac{g_s}{3(2\pi)^{3}}\int_{0}^{k_{f_{j}}} \frac{k^2}{\sqrt{k^2 + m^2_{j}}} \, d^{3}k
\left[f_k(T)+\bar{f}_k(T)\right].
\end{eqnarray}
Where 
\begin{equation}
\label{eq:ffp}
f_k(T)=\frac{1}{e^{[E_i(k)-\mu_i^*]/T}+1}
\end{equation}
and
\begin{equation}
\label{eq:ffap}
\bar f_k(T) = \frac{1}{e^{[E_i(k)+\mu_i^*]/T}+1}
\end{equation}
are the Fermi distribution functions for particle and anti-particle at finite temperature T, respectively. Both $f_k(T)$ and $\bar{f}_k(T)$ appear in the integrals. This reflects the thermal population of anti-particles, which grows more important as temperature approaches the nucleon effective mass $M^*_{\rm nucl.}$. For newly born neutron stars ($T \sim 10$--$80$ MeV) \cite{AA467-395-2007}, $M^*_{\rm nucl.} \gg T$. So anti-particle contributions stay exponentially suppressed, but they are still non-negligible for precision calculations. The effective nucleon mass $M^*_{\rm nucl.}$, defined as $M_{nucl.}^\ast = M_N - g_\sigma \sigma_0 \mp g_\delta \delta_0$, decreases with increasing density as the scalar field $\sigma_0$ grows. This reflects the partial restoration of chiral symmetry in dense matter. It is physically important because a smaller $M^*_{\rm nucl.}$ softens the nuclear EoS at intermediate densities and modifies the stellar mass-radius curve. This density-dependent mass also enters the axion emission rate through ${m^*}^2$ in Eq.~(\ref{eq:etheta}). So nuclear medium effects are transmitted directly to the axionic cooling channel.
\\
\\
The $\sigma-$, $\omega-$, $\rho-$ and $\delta-$ are the meson fields for iso-scalar scalar $\sigma-$meson, iso-scalar vector $\omega-$meson, iso-vector vector $\rho-$meson and iso-vector scalar $\delta-$mesons, respectively. The $g_{\sigma}$, $g_{\omega}$, $g_{\rho}$ and $g_{\delta}$ are the coupling constants and $g_s$ is the internal degrees of freedom.
The corresponding masses are $m_{\sigma}$, $m_{\omega}$, $m_{\rho}$ and $m_{\delta}$ for $\sigma$, $\omega$, $\rho$ and $\delta$ mesons. %The effective mass of the nucleon is defined as 
 %\begin{equation}
 %\label{eq:emass}
%M_{nucl.}^\ast = M_N - g_\sigma \sigma_0 \mp g_\delta \delta_0 .
%\end{equation}
%
 The electromagnetic interaction is incorporated through the Coulomb coupling constant $e^{2}/4\pi$. From the  energy-momentum tensor, one can find the expression for energy density  Eq.~(\ref{eq:eden})  and pressure  Eq. (\ref{eq:press}). These equations are solved self-consistently through an iterative numerical procedure and detail numerical procedure is available in Refs. 
\cite{PRC97-045806-2018,NPA966-197-2017}.\\\\
\subsection{Quarkyonic Model} \label{b}
At sufficiently large baryonic density ($n_B \geq 2.5 n_0$, where $n_0$ is the nuclear saturation density), the quark substructure of baryons starts to be relevant. The quark medium becomes important with increase of baryon density $n_B$. In such condition, quarks fill the low-momentum states, while the nucleons stay clustered near the Fermi surface \cite{JCAP2025-056-2025,PRL122-122701-2019,PRD102-023021-2020}. The consequence of this scenario is a sharp stiffening of the EoS near the transition density $n_t$. Considering, the baryon-number conservation, charge neutrality, and $\beta$-equilibrium together for the part of the supernova remnant, which is neutron star matter \cite{JCAP2025-056-2025}, we get
\begin{eqnarray}
n_B &=& n_n + n_p + \frac{n_u+n_d}{3}
\nonumber\\
&=& \frac{g_s}{6\pi^2}\bigg[(k_{f_{n}}^3-k_{0_{n}}^3)+(k_{f_{p}}^3-k_{0_{p}}^3)+\frac{(k_{f_{u}}^3+k_{f_{d}}^3)}{3}\bigg],
\end{eqnarray}
The charge neutrality condition gives the relation:
\begin{eqnarray}
n_p + \frac{2n_{u}}{3} -  \frac{n_{d}}{3}= n_{e^{-}} + n_{\mu}.
\end{eqnarray}
Following \cite{JCAP2025-056-2025}, the lower momentum bounds for the nucleons are tied to the transition momentum $k_{\rm Ft}$ through:
\begin{eqnarray}
 k_{0(n,p)} &=& (k_{f_{(n,p)}}-k_{t_{(n,p)}})\bigg[1+ \frac{\Lambda_{\rm cs}^2}{k_{f_{(n,p)}}k_{t_{(n,p)}}}\bigg].
\end{eqnarray}
For neutrons, protons, and quarks to remain in chemical balance, we need:
\begin{eqnarray}\label{ebnq}
\mu_n &=& \mu_u + 2\mu_d, \\
\mu_p &=& 2\mu_u + \mu_d.
\end{eqnarray}
Beta-equilibrium, driven by the weak interaction, further requires:
\begin{eqnarray} \label{qnbe}
\mu_{n} &=& \mu_{p} + \mu_{e^{-}}, \nonumber \\
\mu_{\mu} &=& \mu_{e^{-}}.
\end{eqnarray}
At the onset, the light-quark masses are fixed by:
\begin{eqnarray}
m_u &=& \frac{2}{3} \mu_{t_p} - \frac{1}{3} \mu_{t_n}, \nonumber \\
m_d &=& \frac{2}{3} \mu_{t_n} - \frac{1}{3} \mu_{t_p}.
\end{eqnarray}
The quarks are treated as free and non-interacting Fermi gas, which leads to the energy density ${\cal{E}}_{QM}$ and pressure $P_{QM}$ at T=0 MeV as \cite{JCAP2025-056-2025,IJMPD35-2650015-2026}:
\begin{eqnarray}
\label{eq:eqm}
{\cal E}_{\rm QM}&=& \sum_{j=u,d}\frac{g_s N_c}{(2\pi)^3}\int_0^{k_{f_{j}}}\sqrt{k^2 + m_{j}^2 }\, d^3k,
\end{eqnarray}
\begin{eqnarray}
\label{eq:eqp}
P_{\rm QM} &=& \mu_{u} n_{u} + \mu_{d} n_{d} - \cal E_{\rm QM}.
\end{eqnarray}
\\\\
In the above equations, $n_i$ ($i=n,p,u,d,e^{-},\mu$) represents the number density of neutrons, protons, up quarks, down quarks, electrons, and muons, respectively. $k_{f_i}$ is the Fermi momentum of each species, $k_{0(n,p)}$ is the minimum nucleon momentum reached in the quarkyonic phase, and $k_{t(n,p)}$ is the transition momentum linked to the transition density $n_t$ marking the start of the quarkyonic regime. The $g_s=2$ is the spin and $N_c=3$ is the color degeneracy factors. The $\mu_i$ is the chemical potential of each species $i$. The QCD confinement scale $\Lambda_{\rm cs}$  separate the quark and nucleon momentum ranges.
As with the baryonic sector, the newly formed supernova remnant sits at high temperature. Because of this, the quarkyonic-sector EoS must also be evaluated in the high-temperature limit, which are written in finite temperature T as \cite{ANP16-52-1986}:
\begin{eqnarray}
\label{eq:ftqm}
{\cal E}_{\rm QM}(T)
=\sum_{j=u,d}\frac{g_s N_c}{(2\pi)^3}\int_0^{k_{f_{j}}}\sqrt{k^2 + m_{j}^2 }\, d^3k[f_k(T)+\bar{f}_k(T)]
\end{eqnarray}
\begin{eqnarray}
\label{eq:ftqp}
P_{\rm QM}(T)
= \mu_{u} n_{u} + \mu_{d} n_{d} - {\cal E_{\rm QM}}(T).
\end{eqnarray}
Here, $[f_k(T)+\bar{f}_k(T)]$ represents the combined Fermi distribution of quarks and anti-quarks. We treat quarks as a free, non-interacting Fermi gas. This is motivated by asymptotic freedom: at high momentum transfers, the QCD running coupling $\alpha_s(Q^2)$ decreases logarithmically, and quark-quark interactions weaken. This makes the leading-order free-gas description a well-controlled starting point for the bulk thermodynamics. The chemical equilibrium conditions in Eqs.~(\ref{ebnq}) and~(\ref{qnbe}) keep the quark and nucleon sectors mutually consistent at every density. No net strangeness accumulates, and overall charge neutrality of the star is preserved. Incorporating finite-temperature Fermi distributions for quarks is essential. The same temperature that excites thermal nucleons in the hadronic sector also populates quark states near the Fermi surface in the quarkyonic core. Neglecting this thermal quark contribution would break thermodynamic consistency across the crossover boundary. It would also artificially underestimate the total pressure at finite~$T$.
% \subsection{Quarkyonic-inspired crossover equation of state}
The quarkyonic equations of state used here are built on a smooth crossover scheme, adapting the interpolated-EoS prescription put forward by Masuda \emph{et al.}~\cite{PTEP2013-073D01-2013} and later applied by Han \emph{et al.}~\cite{PRD100-103022-2019}.
Rather than imposing a sharp hybrid (Maxwell or Gibbs) matching \cite{PRC74-055803-2006,PRC75-035808-2007},
this method produces a gradual
transition between hadronic and quark (or quarkyonic-inspired) degrees of
freedom. Specifically, the pressure is smoothly
blended between the hadronic and quarkyonic-inspired branches via a continuous
switching function, whose finite width $\Gamma$
sets how broad the crossover region is. Such a construction sidesteps unphysical
jumps and keeps the numerics well-behaved, all while retaining the key
high-density stiffening that quarkyonic dynamics produces.
\\\\
Within this scheme, the transition density $n_t$ sets a characteristic
scale for the onset of quarkyonic degrees of freedom, while the crossover itself is spread over a finite density window governed by the width parameter $\Gamma$.
Because of this, the EoS need not match the purely
nucleonic EoS exactly up to one sharp density cutoff.
% \subsubsection*{Pressure interpolation}
Following~\cite{PRD100-103022-2019}, the crossover is built by interpolating the pressure in baryon
density space:
\begin{equation}
P(n_B) = P_{\rm NM}(n_B)\,f_{-}(n_B) + P_{\rm QM}(n_B)\,f_{+}(n_B).
\label{eq:Pinterp}
\end{equation}
Here $P_{\rm NM}(n_B)$ and $P_{\rm QM}(n_B)$ are, as introduced above [Eqs.  (\ref{eq:press})  and  (\ref{eq:ftqp})], the purely nucleonic and
quarkyonic-inspired pressures. The smooth switching functions obey
\begin{equation}
f_{-}(n_B) + f_{+}(n_B) = 1,
\qquad
f_{\pm}(n_B) =
\frac{1}{2}
\left[
1 \pm \tanh\!\left(\frac{n_B-n_t}{\Gamma}\right)
\right].
\label{eq:weights}
\end{equation}
The $\Gamma$  is treated here as a phenomenological input chosen to keep the interpolation smooth and numerically well-behaved. Throughout this work it is held fixed at $\Gamma = 0.05\,{\rm fm}^{-3}$ for every transition density $n_t$ studied.
% \subsubsection*{Thermodynamically consistent energy density}
As noted in Refs.~\cite{PTEP2013-073D01-2013,PRD100-103022-2019}, interpolating the pressure alone is not enough — the energy density must be reconstructed in a way that stays thermodynamically consistent. In the crossover region it therefore takes the form
\begin{equation}
{\cal{E}}(n_B)
=
{\cal{E}}_{\rm NM}(n_B)\,f_{-}(n_B)
+
{\cal{E}}_{\rm QM}(n_B)\,f_{+}(n_B)
+
\Delta{\cal{E}}(n_B),
\label{eq:energy}
\end{equation}
matching Eq.~(21) of Ref.~\cite{PRD100-103022-2019}. The extra piece
$\Delta{\cal{E}}(n_B)$ comes from the fact that the switching function itself depends on density and is given by
\begin{equation}
\Delta{\cal{E}}(n_B)
=
n_B \int_{n_t}^{n_B}
dn'\,
\frac{{\cal{E}}_{\rm NM}(n_B') - {\cal{E}}_{\rm QM}(n_B')}
{n_B'}\, g(n_B'),
\label{eq:deltaeps}
\end{equation}
with
\begin{equation}
g(n_B') = \frac{d f_{+}(n_B')}{dn_B'}
      = \frac{1}{2\Gamma}
        \mathrm{sech}^2\!\left(\frac{n_B'-n_t}{\Gamma}\right).
\end{equation}
This term is not an arbitrary add-on — it follows directly from imposing the
thermodynamic relation
\begin{equation}
P = n_B^2 \frac{\partial}{\partial n_B}\left(\frac{{\cal{E}}}{n_B}\right),
\end{equation}
Since the interpolation reshapes how baryon density $n_B$, energy
density ${\cal{E}}$, and pressure $P$ map onto one another, two EoS curves that
look nearly indistinguishable in the $P(n_B)$ picture can appear visibly separated once plotted in
$P({\cal{E}})$. So the seeming split between the EoS curves at low energy
density is really just a signature of the smooth-crossover construction, not evidence of any change to the underlying nucleonic E-RMF physics. We
stress that the nucleonic parameter sets (G3) remain unchanged throughout and are never refit for different values of $n_t$. An important consistency check of the crossover construction is the speed of sound $c_s^2 = \partial P/\partial \mathcal{E}$. A physically admissible EoS must satisfy $0 \leq c_s^2 \leq 1$ (in natural units) at all densities. Violating the upper bound would imply superluminal propagation, making the EoS unphysical. The smooth tanh-based switching function in Eq.~(\ref{eq:weights}) lets $c_s^2$ evolve continuously. It moves from the hadronic value at low densities to the quarkyonic-dominated value at high densities. It avoids the sharp discontinuities of Maxwell constructions, which can produce unphysical dips in $c_s^2$. The crossover width $\Gamma = 0.05\,{\rm fm}^{-3}$ is chosen carefully. It is narrow enough that quarkyonic stiffening occurs within the density range probed by $\sim 2\,M_\odot$ pulsars. It is also broad enough to keep the TOV integration numerically stable across the full range of central densities explored.
%%%%%%%%%%%%%%%%%%%%%%%%%%%%%%%
%%%%%%%%%%%%%%%%%%%%%%%%%%%%%%%%
%%%%%%%%%%%%%%%%%%%%%%%%%%%%%%%%%%%%%%%%%%%%%%%%%%%%%%%%%%%%%%%%
%%%%%%%%%%%%%%%%%%%%%%%%%%%%%%%%%%%%%%%%%%%%%%%%%%%%%%%%%%%
\subsection{Axion Model} \label{c}
The energy and pressure of the NS due to the presence of axions is derived from the interaction of axion fields with the nucleons. The interaction part of the Lagrangian density is defined as: \cite{PRL53-1198-1984},
\begin{equation}
\label{eq:lag}
{\cal L_{\theta}} = ig_{\theta}{\bar{\psi}}\gamma^5\psi\Theta.
\end{equation}
Here, the strength of interaction axion with nucleon $g_{\theta}=\frac{C_{\theta}m}{F}$ with $C_{\theta}$ is the axial vector renormalization constant and $m$ is the mass of the nucleon. Our aim is to study the interaction of axion with the nucleons inside the NS matter along with the emission of energy loss rate of axion. Considering the one pion exchange potential for the nucleon-nucleon collisions in the $\pi-n$ pseudo vector coupling derived from the $p-$wave the expression for the loss of energy rate in the process $n+n\rightarrow n+n+\theta$ is given as \cite{PRL53-1198-1984}:
\begin{equation}
\label{eq:etheta}
{\cal E_{\theta}} = \frac{31}{1890\pi}g_{\theta}^2\left(\frac{f}{m_{\pi}}\right)^4{m^*}^2p_F(n)F(x)\left(k_{\beta}T\right)^6,
%\end{eqnarray}
\end{equation}
with $x=\frac{m_{\pi}}{2p_F(n)}$ and $p_F(n)$ is the nucleon Fermi momentum. Due to the very small axion mass ($m_{\theta}\sim 10^{-5}$ eV), it is neglected in the calculations. The values of $f$, $F(x)$ and $k_{\beta}$ are taken as 1. The second process  $e^-+(Z,A)\rightarrow e^-+(Z,A)+\theta$ takes place on the crust of the NS, which is composed of  a lattice of ion and degenerate electrons gas. The interaction Lagrangian density is therefore written as
\begin{equation}
\label{eq:lag}
{\mathcal L_{\theta}^e} = ig_{\theta_e}{\bar{\psi_e}}\gamma^5\psi_e\Theta,
\end{equation}
where the coupling strength  $g_{\theta_e}=\frac{m_e}{F}$ and $m_e$ is the mass of the electron.
Then the energy loss rate due to emission of axion in the electron-ion collision process is written as 
\begin{eqnarray}
\label{eq:lag}
{\mathcal E_{\theta}^e} = \frac{\pi^2}{120}\frac{Z^2\alpha}{A} g_{\theta_e}^2\frac{n_b(k_{\beta}T)^4}{p_F(e)^2} [2ln(2\gamma)-1],
\end{eqnarray}
where $\gamma$ is the Lorentz factor of the electron and $p_F(e)$ is it's Fermi momentum. The electromagnetic coupling constant $\alpha=\frac{e^2}{4\pi}=\frac{1}{137}$. \\
The free energy density of the axion at finite temperature T is written as \cite{BTFT925-1-2016,arxiv:2506.1692v2}:
%,UNIVERSE11-2025}.
\begin{eqnarray}
\label{eq:lag}
{\cal F}_{\theta}(T) \approx -\frac{\pi^2T^4}{90}+\frac{m_{\theta}^2T^2}{24} -\frac{m_{\theta}^3T}{12\pi}+\frac{3}{4}\lambda\left(\frac{T^4}{144}-\frac{m_{\theta}T^3}{24\pi}\right)\nonumber\\+{\cal O}(\lambda^2)+......
\end{eqnarray}
The thermal pressure $P(T)$ for a Fermionic system (nucleons and quarks)  is defined in   Eqs.~(\ref{eq:press}) and ~(\ref{eq:eqp}). It is worthy to note that the $P(T)$ suppressed exponentially at low T.  
%in the non-relativistic limit can be written as\cite{TEU69-1990,LAP2019,arxiv:2506.1692v2}:
%\begin{eqnarray}
\label{eq:lag}
%P^{th}(T) =g_fT\left(\frac{mT}{2\pi}\right)^{3/2}exp(-m/T)\right). 
%\end{eqnarray}
As far as the nucleons and quarks are concerned, the stability of the neutron star is due to the nucleons and quarks degeneracy pressure, which are Eqs.~(\ref{eq:press}) and ~(\ref{eq:eqp}). The total equation of state of the axion-admixed quarkyonic star is the sum of three contributions: $\mathcal{E}_{\rm tot} = \mathcal{E}(n_B) + \mathcal{E}_\theta(T)$ and $P_{\rm tot} = P(n_B) + P_\theta(T)$. Here $\mathcal{E}(n_B)$ and $P(n_B)$ are the crossover hadronic-quarkyonic EoS from Eqs.~(\ref{eq:energy}) and~(\ref{eq:Pinterp}). $P_\theta = -\mathcal{F}_\theta$ is the axionic pressure derived from Sec.~(\ref{eq:lag}). This combined EoS is fed into the TOV equations \cite{PR55-364-1939,PR55-518-1939}. This gives the stellar mass $M$, radius $R$, compactness $C = M/R$, and dimensionless tidal deformability $\Lambda$ as functions of central density and temperature. The axion contribution modifies both the pressure support and the energy content of the star. So it's effect on the mass-radius relation and on $\Lambda$ is self-consistent, not a post hoc correction. This distinction matters most at the high temperatures of the early hot neutron star phase.
%\cite{LAP2019,arxiv:2506.1692v2}:
%\begin{eqnarray}
%\label{eq:lag}
%P^{D}(T) =g_fT\left(\frac{mT}{2\pi}\right)^{3/2}exp(-m/T)\right). 
%\end{eqnarray}
Similarly, the thermal energy density of the supernova remnant is defined in Eqs.~(\ref{eq:eden}) and ~(\ref{eq:ftqm}). These equations are valid as long as there is no trapping of axions. However, to respect the strong CP problem, it is mandatory to include the pseudo-scalar Goldstone particles in the standard model as it is discussed in the earlier section. Accordingly, we have Eqs.~(\ref{eq:eda}) and ~(\ref{eq:epa})
are the energy density expressions for the core and crust of the neutron star. Also, as it is stated already, the main mechanism
for the axion production is the Bremsstrahlung process due to the nucleon-nucleon collision, which occurs mostly in the interior of the neutron star and we can neglect the 
axion production on the surface due to the collision of 
electrons with the lattice ions. Another reason for the neglect of the axion flux at the surface region is the Kim-Shifman-Vainshtein-Zakharov (KSVZ) model \cite{PRL43-103-1979,NPB166-493-1980}. Unlike the Dine-Fischler-Srednicki-
Zhitnitskii (DFSZ) model \cite{PLB104-199-1981,YF31-497-1980}, in the KSVZ model the axion-electron coupling vanishes at the tree level. Thus, the total thermodynamic quantities of the system are obtained by combining the nucleonic, quark and axionic contributions as
\begin{eqnarray}
\label{eq:eda}
{\cal{E}}(T) &=& {\cal{E}}_{\rm NM}(T) + 
{\cal{E}}_{\rm QM}(T)+{\cal{E}}_{\theta}(T),
\label{eq:etot}
\end{eqnarray}
\begin{eqnarray}
\label{eq:epa}
P(T) &=& P_{\rm NM} (T)+ P_{\rm QM}(T)+P_{\theta}(T).
\label{eq:ptot}
\end{eqnarray}
Where the axion pressure is defined as the negative of the 
free energy density $F_{\theta}(T)$, i.e., $P_{\theta}(T)=-{\cal{F}}_{\theta}(T)$.
\\
\subsection{Tolman--Oppenheimer--Volkoff equations}
Once knowing the thermal pressure $P(T)$ and the thermal energy ${\cal{E}}(T)$ from equations Eqs.~(\ref{eq:etot}) and ~(\ref{eq:ptot}), the mass and radius of the neutron star at finite T can be obtained from the Tolman-Oppenheimer-Volkoff equations \cite{PR55-364-1939,PR55-518-1939}. For a given $P(T)$ and ${\cal{E}}(T)$, stellar equilibrium is obtained by integrating the TOV equations \cite{PR55-364-1939,PR55-518-1939}:
\begin{align}
\frac{dP(r)}{dr} &= -\frac{G}{r^2}\,\frac{\big[{\cal E} (r)+P(r)\big]\big[M(r)+4\pi r^3 P(r)\big]}{1-2GM(r)/r}, \\
\frac{dM(r)}{dr} &= 4\pi r^2 {\cal E} (r),
\end{align}
with central conditions $M(0)=0$ and $P(0)=P_c$. For chosen central values (equivalently, central baryon and DM densities) the integration proceeds outward until $P(R)=0$, determining the stellar radius $R$ and gravitational mass $M(R)$.
%%%%%%%%%%%%%%%%%%%%%%%%%%%%%%%
%%%%%%%%%%%%%%%%%%%%%%%%%%%%%%%%
%%%%%%%%%%%%%%%%%%%%%%%%%%%%%%%%%%%%%%%%%%%%%%%%%%%%%%%%%%%%%%%%
%%%%%%%%%%%%%%%%%%%%%%%%%%%%%%%%%%%%%%%%%%%%%%%%%%%%%%%%%%%
\section{Results and Discussions} \label{result}
\begin{figure}
\includegraphics[width=1.0\columnwidth]{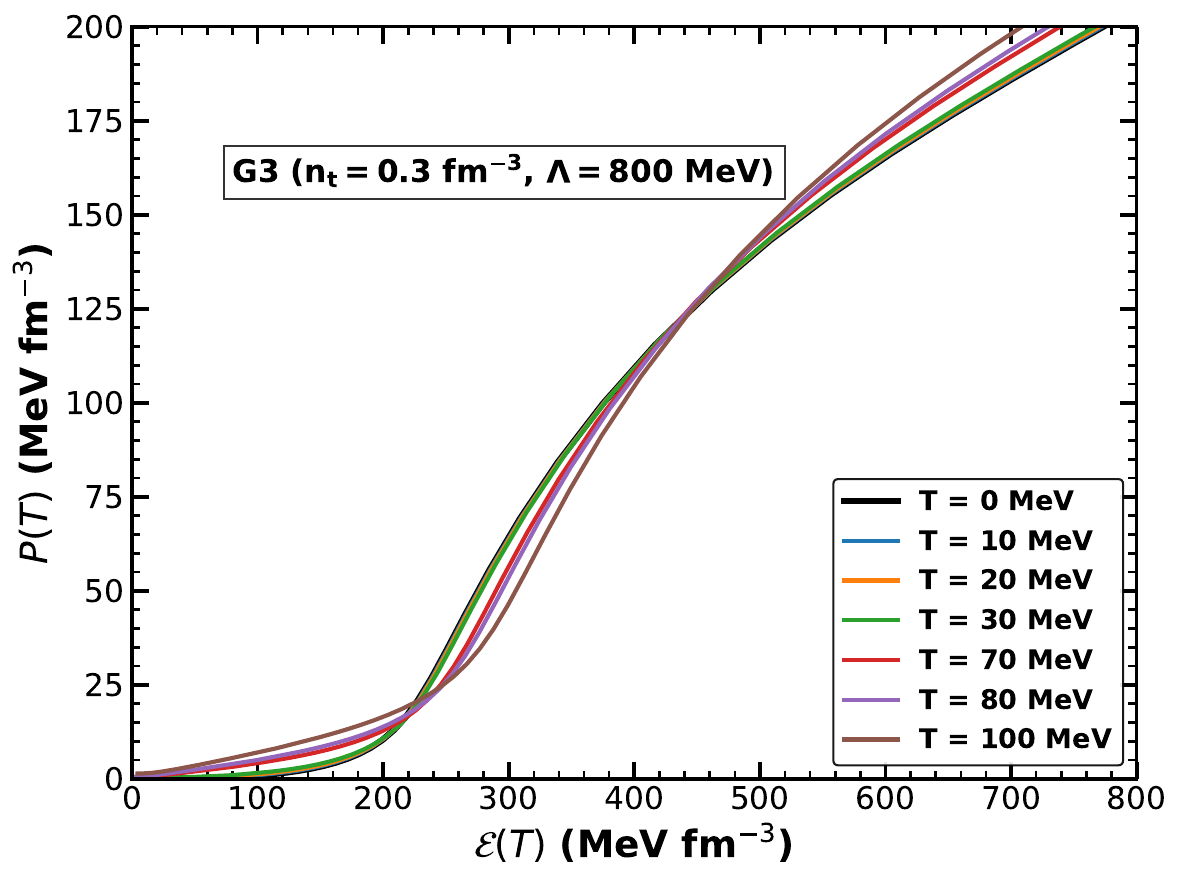}
\caption{Pressure $P(T)$ as a function of energy density ${\cal{E}}(T)$ for the quarkyonic equations of state at a transition density
$n_t = 0.3 fm^{-3}$ and  confinement scales
$\Lambda_{\mathrm{cs}} = 800$~MeV for the G3 parameter set at
different temperatures. }
% \textcolor{red}{Alt text: Line plot of pressure versus energy density for the quarkyonic-axion equation of state, showing multiple curves for temperatures from 0 to 100 MeV; the higher-temperature curves are stiffer near the stellar core and surface, while the T=0 curve is stiffest in the intermediate region.}
\label{fig:eos}
\end{figure}

In our calculations, the transition density from the quarkyonic to the hadronic phase is taken as $n_t=0.3$ fm$^{-3}$ with the QCD confinement scale $\Lambda_{cs}$ = 800 MeV. The parameter $n_t$ characterizes the central density of the crossover region and represents a sharp phase boundary between the quark and baryonic matter, respectively.
We used the recently proposed quarkyonic model of McLerran and Reddy \cite{PRL122-122701-2019} which is further extended by Dey et al \cite{JCAP2025-056-2025} substituting the E-RMF formalism for the  baryonic sector and making a smooth cross-over phase transition from quark to baryonic phase using the prescriptions outline in Refs. \cite{PTEP2013-073D01-2013,PRD100-103022-2019}. The $\beta-$equilibrium along with the charge neutrality is also maintained for the NS matter. We studied the effects of EoS and  mass (M), radius (R), compactness (C) and dimensionless tidal deformability ($\Lambda$) for an isolated non-spinning neutron star that emits axions. The axions are predicted to be a dark matter particle, whose presence is important to satisfy the strong CP violation in the standard model. Thus we incorporated the axion in our calculations and explore it's effects at various temperature of the supernova remnant just after the SN explosion. The mass of the axion particle is extremely small and many literature suggest it's mass $m_{\theta}$ is
$\leq 1$ eV and the axion-nucleon coupling constant $g_{\theta}\sim 10^{-9}$ \cite{PRL128-091102-2022}. Here, we have used these parameters for the axionic EoS, i.e., $m_{\theta}=1.0$ meV (0.001 eV=1 milli electron volt) and $g_{\theta}=1.3\times{10}^{-9}$. The EoSs of the hybrid quarkyonic model are used in the TOV equations to evaluate the observational properties of neutron star, which we considered as a remnant of supernova explosion.
\\
\\
We note that the equations of state shown in Figure~\ref{fig:eos} are presented in a pressure--energy density representation, $P({\cal{E}})$. In the crossover framework adopted here, the interpolation occurs in baryon density space, $P(n_B)$, while the energy density is obtained from Eq.~(\ref{eq:energy}). As a result, equations of state that are nearly identical in the $P(n_B)$ representation may appear separated when expressed as $P({\cal{E}})$, particularly in the vicinity of the crossover region. This behavior is a generic feature of smooth crossover constructions and does not imply a modification of the underlying nucleonic E-RMF dynamics.
The figure also shows that the quarkyonic crossover remains smooth in all cases: there is no abrupt plateau or discontinuity in $P({\cal{E}})$, and the slopes remain positive.
%%%%%%%%%%%%%%%%%%%%%%%%%%%%%%%%%%%%%%
%%%%%%%%%%%%%%%%%%%%%%%%
Figure~\ref{fig:eos} shows the pressure as a function of energy density of the neutron star including axions at temperature T= 0, 10, 20, 30, 70, 80, 100 MeV. The transition density $n_t=0.3$ fm$^{-3}$ and QCD confinement scale is assumed to be $\Lambda{cs}=800$ MeV for the quark sector and the G3 parameter set \cite{PRC97-045806-2018} is used for the baryonic phase of the EoSs. The values of the G3 parameters and the corresponding properties of the nuclear matter at saturation are listed in Table I. It is worth mentioning that the G3 parameter set not only reproduces the nuclear matter properties well, but also predicts the finite nuclear properties throughout the mass table including the $\beta-$stable and away from the $\beta-$stability line. Also, the nuclear matter saturation properties in the G3 set is well within the experimental/empirical range.  The experimental/empirical values are given just below the model's predictions \cite{PRC97-045806-2018,NPA966-197-2017} in Table I. Within the quark and baryonic medium, the axion which has a very light mass is trapped inside the supernova remnant at finite temperature. The mean free path $\lambda_{\theta}$ of the axion is more than the radius of the remnant and plays a pivotal role in the cooling process of the SN remnant to form a neutron star.  The copious emission of the axions flux $\cal{E}_{\theta}$ makes the NS cooler much faster than the emission of neutrinos due to the $\beta-$decays to maintain the thermodynamical equilibrium and charge neutrality, i.e., for stars in which the strongly interacting particles are
baryons. The composition is determined by the requirements
of charge neutrality and $\beta-$equilibrium conditions under 
weak decay processes $B_1\rightarrow B_2+l+\bar{\nu}$ and 
$B_2+l\rightarrow B_1+\nu_l$ \cite{PRC74-055803-2006,PRC75-035808-2007}. Using this Gibbs criteria, we construct the complete EoS taking into account the quarks, baryons and axions in the NS medium at different temperatures and presented in  Figure~\ref{fig:eos}. 

The black solid line represents the $P(T)-{\cal{E}}(T)$ curve at T = 0. With respect to this curve, the other lines for different T are shown. From the figure, it is noticed that the line for T = 0 behaves very differently in different parts of the stellar object. It is one of the stiffest lines in the intermediate region, while it shows softness near the core and the surface area of the neutron star. On the other hand, unlike the T = 0 MeV one, the $P(T)-{\cal{E}}(T)$ line for T=100 MeV is the stiffest curve near the core and the surface regions. All other intermediate curves are distributed within this said limit as shown in the figure. Accordingly, the masses and radii are obtained, which are discussed subsequently.  
\\\\
%%%%%%%%%%%%%%%%%%%%%%%%
\begin{figure}
\includegraphics[width=1.0\columnwidth]{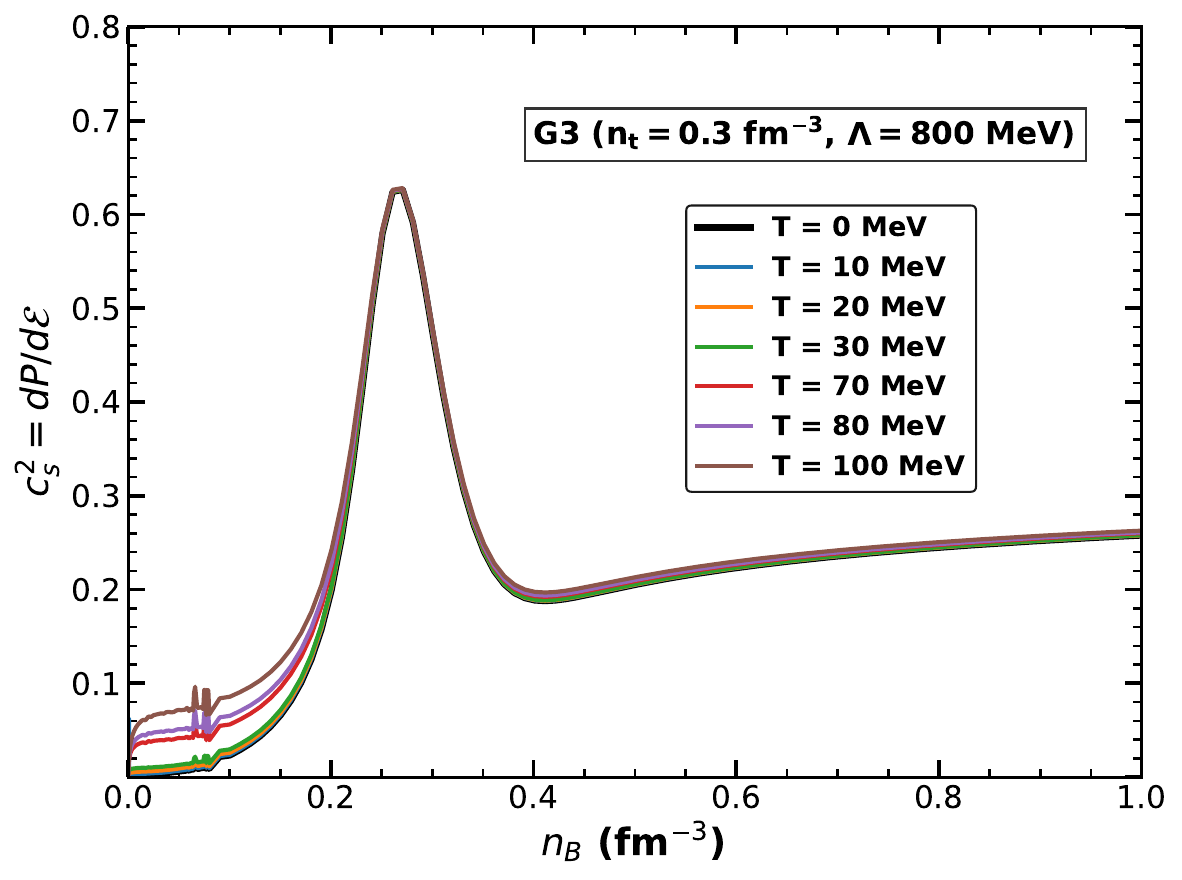}
\caption{The velocity of sound $c_s^2=\frac{dP{(\cal E})}{d\cal E}$ as a function of quark and baryon density $n_B$ at different temperature T for the G3 parameter set. The transition density $n_t=0.3$ fm$^{-3}$ and QCD confinement scale $\Lambda_{cs}=800$ MeV are used in the calculations. }
% \textcolor{red}{Alt text: Line plot of squared sound speed versus baryon density for temperatures from 0 to 100 MeV, showing a kink near the hadron-quark crossover density and values that remain below the causal limit of 1 at all densities and temperatures.}}
\label{fig:cs}
\end{figure}
%%%%%%%%%%%%%%%%%%%%%%%%%%%%%%%%%%%%%%%%%%%%%%%%%%
%%%%%%%%%%%%%%%%%%%%%%%%%%%%%%%%%%%%%%%%%%%%%%%%%%%%%%
\begin{table*}[t]
\centering
\caption{The G3 parameter set and the corresponding nuclear matter properties at saturation density. The nucleon mass is fixed at $M=939$ MeV. The dimension of $k_3$ is fm$^{-1}$, while all other coupling constants are dimensionless. Empirical constraints are taken from Refs.~\cite{Bethe_1971,reed21,shlomo06,colo08,garg18,Colo_2014,Zimmerman_2020,Stone_2014,Pearson_2010,TLi_2010}.}
\label{tab:G3}

%%%%%%%%%%%%%%%%%%%%%%%%%%%%%%%%%%%%%%%%%%%%%%%%%%%%%%%%%%%%%%%%%%%%%%
% Model Parameters
%%%%%%%%%%%%%%%%%%%%%%%%%%%%%%%%%%%%%%%%%%%%%%%%%%%%%%%%%%%%%%%%%%%%%%

\begin{tabular*}{\textwidth}{@{\extracolsep{\fill}}cccccccc@{}}
\hline
\multicolumn{8}{c}{\textbf{Model Parameters}}\\
\hline

$m_{\sigma}/M$ &
$m_{\omega}/M$ &
$m_{\rho}/M$ &
$m_{\delta}/M$ &
$g_{\sigma}/4\pi$ &
$g_{\omega}/4\pi$ &
$g_{\rho}/4\pi$ &
$g_{\delta}/4\pi$\\

0.559 &
0.832 &
0.820 &
1.043 &
0.782 &
0.923 &
0.962 &
0.160\\

\hline

$k_3$ &
$k_4$ &
$\zeta_0$ &
$\eta_1$ &
$\eta_2$ &
$\eta_\rho$ &
$\Lambda_\omega$ &
$\alpha_1$\\

2.606 &
1.694 &
1.010 &
0.424 &
0.114 &
0.645 &
0.038 &
2.000\\

\hline

$\alpha_2$ &
$f_\omega/4$ &
$f_\rho/4$ &
$\beta_\sigma$ &
$\beta_\omega$ &
&
&
\\

$-1.468$ &
0.220 &
1.239 &
$-0.087$ &
$-0.484$ &
&
&
\\
\end{tabular*}

%%%%%%%%%%%%%%%%%%%%%%%%%%%%%%%%%%%%%%%%%%%%%%%%%%%%%%%%%%%%%%%%%%%%%%
% Nuclear Matter Properties
%%%%%%%%%%%%%%%%%%%%%%%%%%%%%%%%%%%%%%%%%%%%%%%%%%%%%%%%%%%%%%%%%%%%%%

\begin{tabular*}{\textwidth}{@{\extracolsep{\fill}}lcccccccc@{}}
\hline
\hline
\multicolumn{9}{c}{
\parbox{0.95\textwidth}{
\centering
\textbf{Nuclear Matter Properties at Saturation}\\
\noindent The nuclear matter saturation density $n_0$ is in fm$^{-3}$, while
$E/A$, $J$, $L$, $K_{\infty}$, $K_{\rm sym}$, $Q_{\rm sym}$,
$K_{\rm asym}$, $Q_0$, $K_{\tau}$, $K_{\rm sat2}$ and $M_0$
are in MeV. The symbols have their usual meanings.}
}\\
%{The symbols have their usual meanings.}\\
\hline
\hline

Source &
$n_0$ &
$E/A$ &
$J$ &
$L$ &
$K_\infty$ &
$K_{\rm sym}$ &
$Q_{\rm sym}$ &
$K_{\rm asy}$\\

\hline

G3 &
0.148 &
$-16.024$ &
31.842 &
49.317 &
243.97 &
$-106.07$ &
915.47 &
$-401.98$\\

Empirical &
0.148--0.185 \cite{Bethe_1971} &
$-(15.0$--$17.0)$ \cite{Bethe_1971} &
33.4--42.8 \cite{reed21} &
69--143 \cite{reed21} &
220--260 \cite{shlomo06,colo08,garg18,Colo_2014} &
$-(174$--$31)$ \cite{Zimmerman_2020} &
--- &
---\\

\hline

Source &
$Q_0$ &
$K_{\tau}$ &
$K_{\rm sat2}$ &
$M_0$ &
&
&
&
\\

\hline

G3 &
$-466.61$ &
$-307.65$ &
$-307.65$ &
2460.98 &
&
&
&
\\

Empirical &
--- &
$-(840$--$350)$ \cite{Stone_2014,Pearson_2010,TLi_2010} &
--- &
--- &
&
&
&
\\

\hline
\end{tabular*}

\end{table*}
%%%%%%%%%%%%%%%%%%%%%%%%%%%%%%%%%%%%%%%%%%%%%%%%%%%%%%
\begin{table}[htbp]
\centering
\caption{Maximum-mass and canonical $1.4\,M_\odot$ properties of the zero-temperature Quarkyonic star and the axion-admixed Quarkyonic star at finite temperature $T$ (MeV): gravitational mass $M$, radius $R$, compactness $C=M/R$ and dimensionless tidal deformability $\Lambda$, evaluated at the maximum-mass configuration (subscript ``max'') and at the canonical mass $M=1.4\,M_\odot$.}
\label{tab:quarkyonic_axion_finiteT}
\begin{tabular}{lccccccc}
\toprule
Model & \multicolumn{4}{c}{Maximum-mass configuration} & \multicolumn{3}{c}{Canonical $1.4\,M_\odot$ star} \\
\cmidrule(lr){2-5} \cmidrule(lr){6-8}
 & $M_{\rm max}$ & $R_{\rm max}$ & $C_{\rm max}$ & $\Lambda_{\rm max}$ & $R_{1.4}$ & $C_{1.4}$ & $\Lambda_{1.4}$ \\
 & ($M_\odot$) & (km) & & & (km) & & \\
\midrule
$T=0$ & 2.750 & 14.540 & 0.279 & 16.040 & 13.684 & 0.151 & 1170.982 \\
\midrule
\multicolumn{8}{c}{Quarkyonic + axion} \\
\midrule
$T=10$ & 2.750 & 14.660 & 0.277 & 16.052 & 14.072 & 0.147 & 1333.836 \\
$T=20$ & 2.747 & 14.970 & 0.271 & 17.481 & 15.012 & 0.138 & 1540.043 \\
$T=30$ & 2.744 & 15.320 & 0.264 & 18.036 & 16.177 & 0.128 & 1657.704 \\
$T=50$ & 2.733 & 15.940 & 0.253 & 17.691 & 19.179 & 0.108 & 2125.307 \\
$T=70$ & 2.716 & 16.290 & 0.246 & 17.221 & 24.843 & 0.083 & 8158.404 \\
$T=80$ & 2.702 & 16.320 & 0.244 & 17.045 & 29.281 & 0.071 & 29044.249 \\
$T=100$ & 2.651 & 16.010 & 0.244 & 17.599 & 24.301 & 0.085 & 10763.679 \\
\bottomrule
\end{tabular}
\end{table}
%

%%%%%%%%%%%%%%%%%%%%%%%%
To check the causality, we compute the speed of sound $c_s^2=\frac{dP{(\cal E})}{d\cal E}$ for different temperatures T, which are shown in  Fig.~\ref{fig:cs}. For all the T, the values of  $c_s^2$ lie well within the causality limit and gives a green signal for further calculations
of the physical observables, such as $M-R$ profiles, emissivity of the axion flux ${\cal{E}}_{\theta}$ and tidal deformability $\Lambda$ of the neutron star which is within the supernova remnant. A further inspection of the Fig.~\ref{fig:cs}, reveals that a kink like structure appears in the $c_s$ graph at $n_B\sim 0.06-0.08$ fm$^{-3}$ for all T. This reflects the phase transition point after the smoothing procedure in the crossover region. Above this transition density, $c_s^2$ rises monotonically and reaches values $\sim 0.5$--$0.7$ at the highest core densities. This is consistent with perturbative QCD expectations that $c_s^2$ approaches $1/3$ from above asymptotically. The causal bound $c_s^2 \leq 1$ holds at all temperatures and densities. This confirms the physical self-consistency of the crossover construction. The sound-speed profiles are only weakly temperature-dependent at $n_B \lesssim 0.3$ fm$^{-3}$. They diverge progressively at supranuclear densities, where the thermal pressure of the quark sector grows significant. This temperature sensitivity of $c_s^2$ at high density is an important diagnostic tool. It governs how the maximum mass and compactness of the star respond to the finite-temperature EoS. Additionally, it sets the scale over which axion production, which also peaks at high density, is influenced by thermal corrections to the EoS.   
\begin{figure}
\includegraphics[width=1.0\columnwidth]{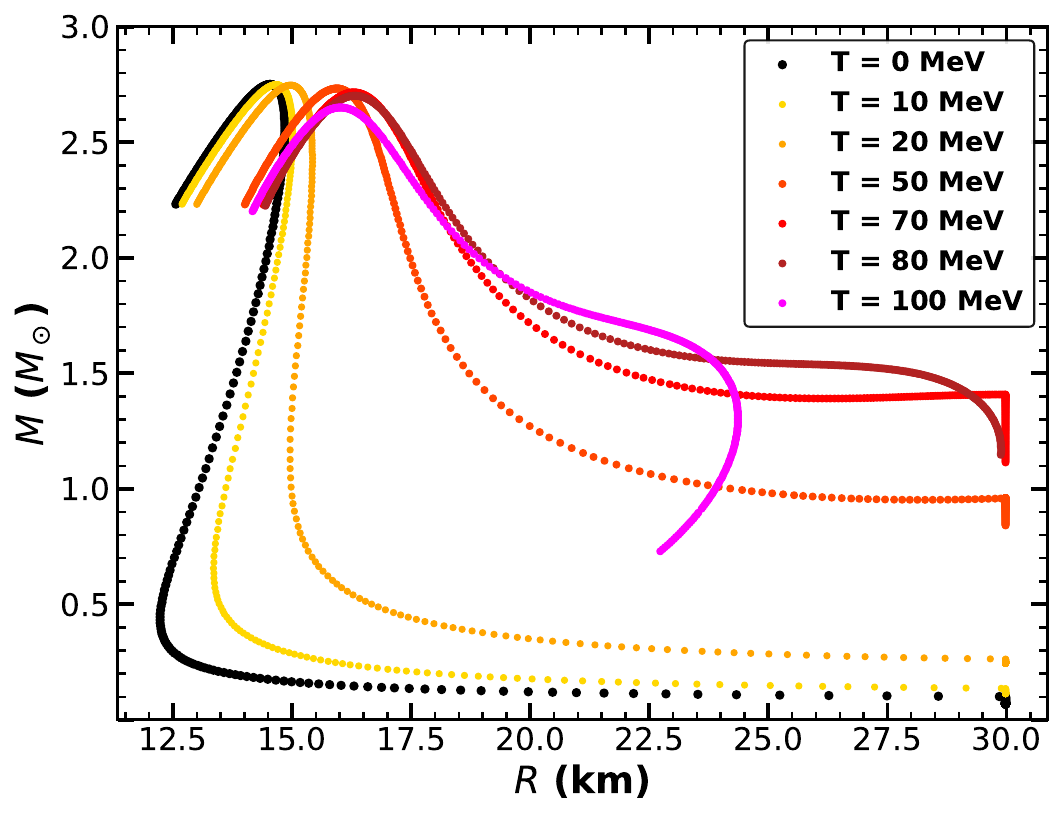}
\caption{The M-R relations for the EoSs shown in Fig. \ref{fig:eos} for different T.The transition density $n_t$=0.3 fm$^{-3}$ and QCD confinement scale $\Lambda_{cs}=800$ MeV is fixed in the calculations.}
% \textcolor{red}{Alt text: Mass-radius curves for the axion-admixed quarkyonic neutron star at temperatures from 0 to 100 MeV, showing the maximum mass staying nearly constant near 2.65-2.75 solar masses while the radius increases substantially with temperature.}}
\label{fig:mr}
\end{figure}
%%%%%%%%%%%%%%%%%%%%%%%%%%
\\\\
The $M-R$ relations for different T are shown in Fig.~\ref{fig:mr}. The maximum masses of the NS remain almost constant with a minimal variation ($M_{max}=2.750$ to 2.651$M_{\odot}$ with T=10 to 100 MeV) for all the temperatures. However, the radius of the NS stretches substantially from 14.66 to 16.32 km with a change of $\sim $ 2 km in the radius. As a result, the compactness parameter $C_{max}=\frac{M_{max}}{R_{max}}$ and dimensionless tidal deformability $\Lambda=\frac{2k_2}{3}\left[\frac{R}{M_{max}G}\right]^5$ 
for a neutron star are also change sizably. The maximum mass $M_{max}$, radius $R_{max}$ and $\Lambda_{max}$ of an isolated static neutron star for different temperatures are tabulated in Table II. The values of $R$, $C$ and $\Lambda$ for a canonical star ($1.4 M_{\odot}$) are also given in the same table. It is interesting to see from Fig.~\ref{fig:mr} and Table II that the NS expands a lot similar to the normal object. However, this expansion of the star limit to some temperature and beyond that there is no more expansion. $M_{\rm max}$ stays nearly constant across the full temperature range. This shows that the maximum mass is controlled by the stiffness of the quarkyonic EoS at high density, which is relatively insensitive to thermal corrections. The radius, by contrast, is inflated substantially by thermal pressure at intermediate densities, where the nucleonic EoS dominates \cite{AA467-395-2007}. The maximum mass $M_{\rm max} \gtrsim 2.65\,M_\odot$ at all temperatures comfortably exceeds the observational lower bound. This bound comes from the heaviest known pulsars, such as PSR~J0952$-$0607 ($M = 2.35\pm0.17\,M_\odot$) \cite{ApJL934-L17-2022}. This confirms that the quarkyonic EoS supports massive compact stars across the full temperature range studied. The canonical radius, however, increases dramatically: from $R_{1.4} = 13.68$ km at $T=0$ to $R_{1.4} = 29.28$ km at $T=80$ MeV. This reflects the amplified thermal sensitivity of lower-mass configurations. These probe the softer, nucleon-dominated part of the EoS, where thermal pressure is a relatively larger fraction of the total. This large radius inflation at finite temperature has a direct observational implication. The tidal deformability $\Lambda_{1.4} \propto R^5$ is enormously enhanced relative to it's cold-star value. So any gravitational-wave signal from a hot neutron star remnant would carry a very different tidal signature than a mature cold star \cite{PRL119-161101-2017}.
\\\\
%%%%%%%%%%%%%%%%%%%%%%%%%%%%%%%%%%%%%%%%%%%%%
%%%%%%%%%%%%%%%%%%%%%%%%%%%%%%%%%%%%%%%%%%%%%%
\begin{figure}
\includegraphics[width=1.0\columnwidth]{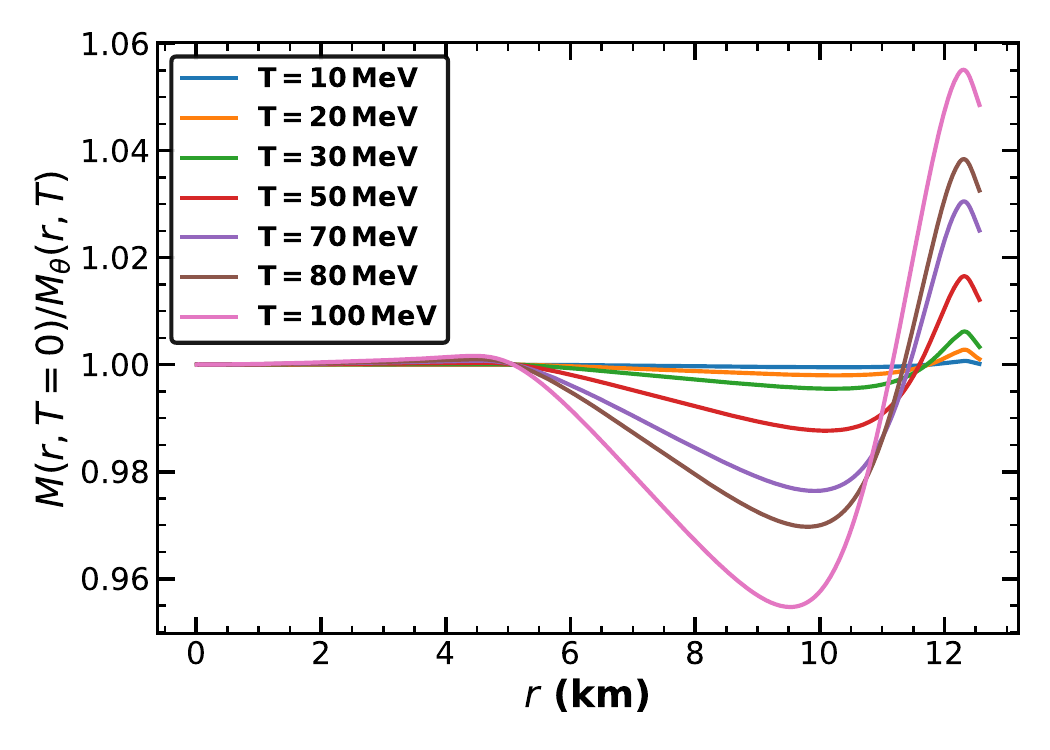
}
\caption{The ratio of the neutron star mass without axion at temperature T=0 MeV to the mass of the NS at T=T including axion contributions $MRatio=\frac{M{(r,T=0)}}{M_{\theta}(r,T)}$. Here, M(r,T=0) is the star mass without axion and $M_{\theta}$ = neutron star mass including the contribution of axions at finite temperature T.} 
% \textcolor{red}{Alt text: Plot of the mass ratio between the neutron star without axions at zero temperature and the axion-admixed star at finite temperature, as a function of radius, showing the deviation growing with temperature and peaking near 9-11 km from the center.}
\label{fig:mrt}
\end{figure}
\\
%%%%%%%%%%%%%%%%%%%%%%%%
%%%%%%%%%%%%%%%%%%%%%%%%
In Fig.~\ref{fig:mrt}, we plot the mass ratio $MRatio=\frac{M{(r,T=0)}}{M{(r,T=T)}}$ of  NS with and without axions' contributions. Here $M{(r,T=0)}$ is the mass of the neutron star without axions at T = 0 MeV and $M{(r,T)}$ is the mass including the baryons and axions' contributions at temperature T=T.  It is interesting to see that the contribution of axion below 10 MeV of the NS is extremely weak. However, the effects are substantial with the increase of temperature. For example, the blue line goes almost straight (line with T=10 MeV), i.e., the $MRatio$ does not deviate much from it's original path. However, it's $MRatio$ goes on deviating more and more with T. It is maximum at T=100 MeV. The second observation is that the emission of axion flux depends very much from the distance of the center of the neutron star. The emission is almost nil starting from the center up to radius $r\sim 5$ km. When $r\geq$ 5 km, the Bremsstrahlung due to nucleon-nucleon collisions taking place with the reaction $n+n\rightarrow n+n+\theta$ and evolution of axion takes place (see Fig.~\ref{fig:mrt}). This process keeps on increasing with temperature and distance from the center of the neutron star with a maximum evolution of axion at $\sim 9-11$ km.  The third observation from Fig.~\ref{fig:mrt}, indicates that the evolved axions start leaving the NS starting from the radius $r\sim 11$ km and total mass of the star reduces gradually and maximum reduction is in the range $r\sim 12-13$ km. In some stage, it so happens that the mass of the NS without axion at T=0 MeV, becomes more than the mass with axion, because more and more energy leaves the star's crust and surface area resulting a decrease in mass. If we estimate a quantitative reduction of mass due to axion production at 100 MeV, with respect to the mass of the NS at T=0 MeV, then the mass reduction is $\sim 5\%$, as shown in Fig.~\ref{fig:mrt}. It also measures the maximum mass reduction at radius $\sim 10$ km, because the $MRatio$ is $\sim 0.95$ at $R \sim 10$ km. The spatial distribution of the axion mass fraction maps where dark matter accumulates in the star. The near-zero axion contribution for $r \lesssim 5$ km reflects the quark-dominated core. There, the baryon density $n_B \gtrsim 3\,n_0$ is so high that nucleon-nucleon Bremsstrahlung is kinematically suppressed relative to the quark processes. So axion production is low. The peak of axion production in the $r \sim 5$--$11$ km shell coincides with the density range where nucleonic degrees of freedom dominate and the nucleon-nucleon collision rate is highest. The decline in $MRatio$ for $r \gtrsim 11$ km shows that the produced axions are not gravitationally bound in the outer layers. They stream outward freely, carrying energy away. This is exactly the additional cooling channel beyond the standard neutrino-dominated scenario \cite{MNRAS363-555-2005,PRD98-103015-2018}. The strong temperature dependence of the mass deficit at large radii implies that axion-driven cooling becomes observationally distinguishable from standard neutrino cooling. This happens during the hottest, earliest phase of neutron star evolution.
\\
%%%%%%%%%%%%%%%%%%%%%%%%
\begin{figure}
\includegraphics[width=1.0\columnwidth]{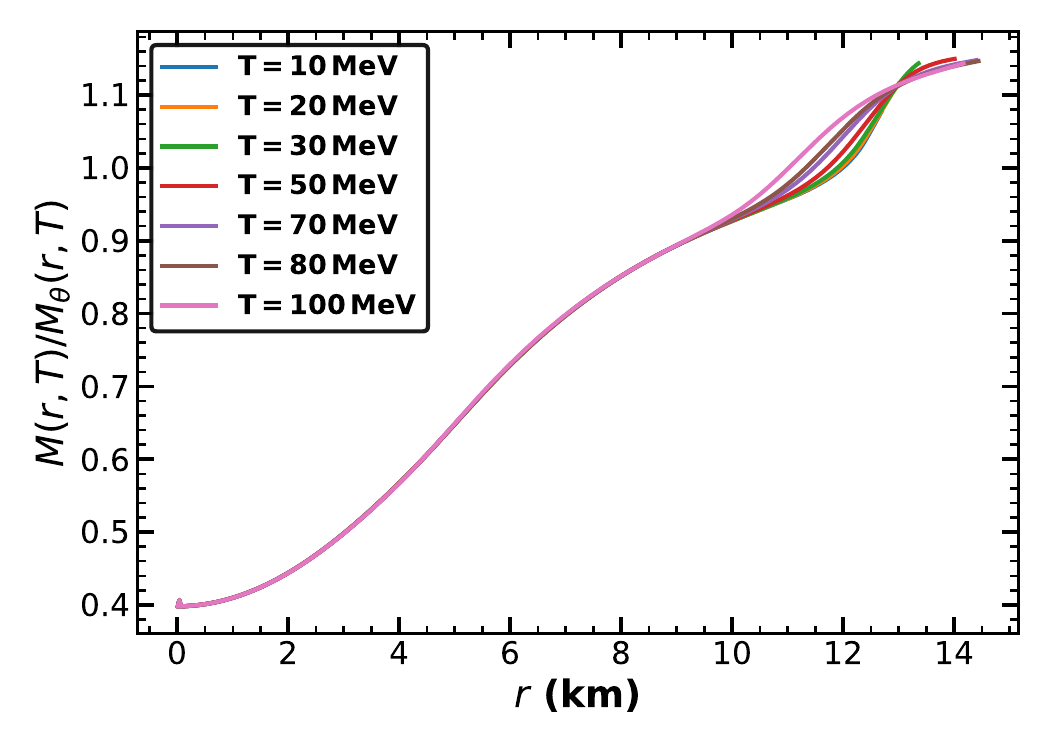}
\caption{The mass ratio  $MRatio(T)=\frac{M{(r,T)}}{M_{\theta}{(r,T)}}$ as a function of distance from the center $r$ at T=10, 20, 30, 50, 70, 80 and 100 MeV. Here, $M{(r,T)}$ = mass of the NS without axion at temperature T and $M_{\theta}{(r,T)}$ is the mass of the NS including axion at temperature T.}
% \textcolor{red}{Alt text: Plot of the ratio of the neutron star mass without axions to the mass including axions at the same finite temperature, as a function of radius, showing a halo-like central depletion with the ratio dropping to about 0.4 near the star's center.}}
\label{fig:mrtt}
\end{figure}
\\
\\
%%%%%%%%%%%%%%%%%%%%%%%%%%%%%%%%%%%%%%%%%%%%%%%%%%
Further, this picture exemplify from the analysis of Fig.~\ref{fig:mrtt}, where we plot the ratio of the pure neutron star (without axiom) at finite temperature (T) to the total mass of the neutron star (including axion) at the corresponding temperature T. When the temperature is taken into account and calculate the mass of the NS without considering the axion contribution, we get almost reverse trend (see Figs. ~\ref{fig:mrt} and ~\ref{fig:mrtt}). The value of  $MRatio(T)=\frac{M{(r,T)}}{M_{\theta}{(r,T)}}$ goes on increasing as we move away from the central region of the star. The trend is nearly similar for all the temperatures as shown in Fig.~\ref{fig:mrtt}. On the other hand, the mass of the NS (without axion) at finite temperature distributed mostly away from the center making a $halo-$type structure in the central region, i.e., $MRatio(T)= 0.4$.  In other words, $M(r,T)$ is $40\%$ smaller than $M_{\theta}(r,T)$ in the center due to the surficial distribution of mass with T. More explicitly, this behavior could be the heavy central mass of the axion star, i.e., 60$\%$ of the mass of the axion is deposited at the center as compared to the finite temperature NS without axion. This implies, although the axions are produced in the intermediate region of the star, mostly it deposits in the quark medium in the center for some time and becomes heavier. The trapping of the axion goes on decreasing with radius and heavily evaporate when the distance is $\sim 11$ km and loses it's mass towards the surface of the axion star. In general, the rate of axion emission is  independent of T, but a careful inspection indicate that within a thickness of 4 km ($r=10$ to 14 km), the emission rate depends on the temperature of the star. It evaporates more in high temperatures as compared to lower T of the medium as shown in Fig.~\ref{fig:mrtt}. The halo-like mass distribution with $MRatio(T) \approx 0.4$ in the central region has a clear physical interpretation. At finite temperature, nucleonic thermal pressure pushes baryonic mass toward larger radii. This reduces the central mass concentration relative to an axion-admixed star at the same temperature. The $60\%$ axion mass concentration in the central region comes directly from axion trapping in the dense quarkyonic core. There, the axion mean free path for re-absorption by nucleons is short, similar to the neutrinosphere in core-collapse supernova physics. This trapped axion population adds to the core energy density and pressure. It stiffens the central EoS and boosts the gravitational mass in the inner $\sim 5$ km relative to a star without axions. Axions gradually evaporate beyond $r \sim 11$ km. The emission is also strongly temperature-sensitive in the $r = 10$--$14$ km shell. This suggests that the ``axion-sphere,'' the radius where axions switch from diffusive to free-streaming, shifts inward at higher temperatures. This behavior could in principle be probed through changes in the neutrino signal from a nearby core-collapse supernova \cite{PRL60-1793-1988,PRD98-103015-2018}.\\
\\
%%%%%%%%%%%%%%%%%%%%%%%%%%%%%%%%%%%%%%%%%%%%%%%%
\begin{figure}
\includegraphics[width=1.0\columnwidth]{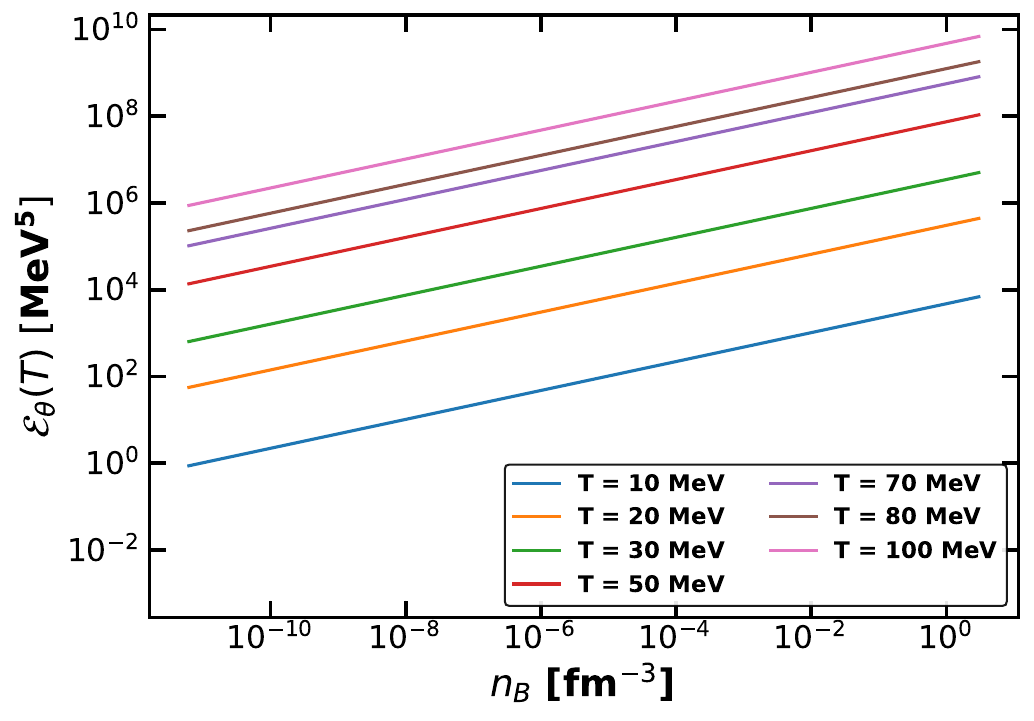}
\caption{The evolution of axion flux ${\cal{E}}_{\theta}(T)$ at different temperature T as a function of baryon density $n_B$. The temperatures are taken as T=10, 20, 30, 50, 70, 80 and 100 MeV.}
% \textcolor{red}{Alt text: Line plot of the axion emission rate as a function of baryon density for temperatures from 10 to 100 MeV, showing emission increasing monotonically with both density and temperature, with the T=100 MeV curve on top and T=10 MeV on the bottom.}}
\label{fig:ea}
\end{figure}
%%%%%%%%%%%%%%%%%%%%%%%%%%%%%%%%%%%%%%%%%%%%%%%%%%
%%%%%%%%%%%%%%%%%%%%%%%%%%%%%%%%%%%%%%%%%%%%%%%%%%
Now the axion emission rate $\cal{E}_{\theta}$(T) is computed from Eq. ~(\ref{eq:etheta}) is presented in Fig.~\ref{fig:ea} at T=10, 20, 30, 50, 70, 80 and 100 MeV as a function of baryon density $n_B$. The energy loss due to axion emission goes on increasing monotonously with baryonic density as well as temperature. This behavior is expected as ${\cal{E}}(T)$ directly proportional to the Fermi momentum and to the $T^6$ of the temperature of the system [see Eq.~(\ref{eq:etheta})]. For lower T, $\cal{E}_{\theta}$(T) corresponds to the lower line in the figure. For example, T=10 MeV corresponds to the most bottom line and T=100 MeV is the top most line. The slope at each point is constant with density. When T is more, movement of the nucleon increases and more number of collision occurs resulting a larger number of axion production. Similarly, when $n_B$ is high, large number of axion present per volume and again number of collision increases causing the production of axion. The $T^6$ scaling of $\mathcal{E}_\theta$ has a striking consequence. Comparing the emissivity at $T = 100$ MeV to that at $T = 10$ MeV gives a ratio of $(100/10)^6 = 10^6$. This explains why axion cooling is negligible in cold mature neutron stars but potentially dominant in the early hot phase \cite{PRL53-1198-1984}. This extreme temperature sensitivity also means the axion luminosity $L_\theta \propto \int \mathcal{E}_\theta \, dV$ rises steeply in the earliest, hottest phase of the star's life. It then falls off rapidly as the star cools. This produces a sharply peaked axion light curve on top of the longer-lived neutrino cooling transient. One observational signature of this burst-like emission is a shortening of the neutrino burst duration from a core-collapse supernova. This is directly constrained by the SN~1987A data \cite{PRL60-1793-1988}. So our finite-temperature axion emissivity calculation is a direct input to that astrophysical bound.\\\\
\begin{figure}
\includegraphics[width=1.0\columnwidth]{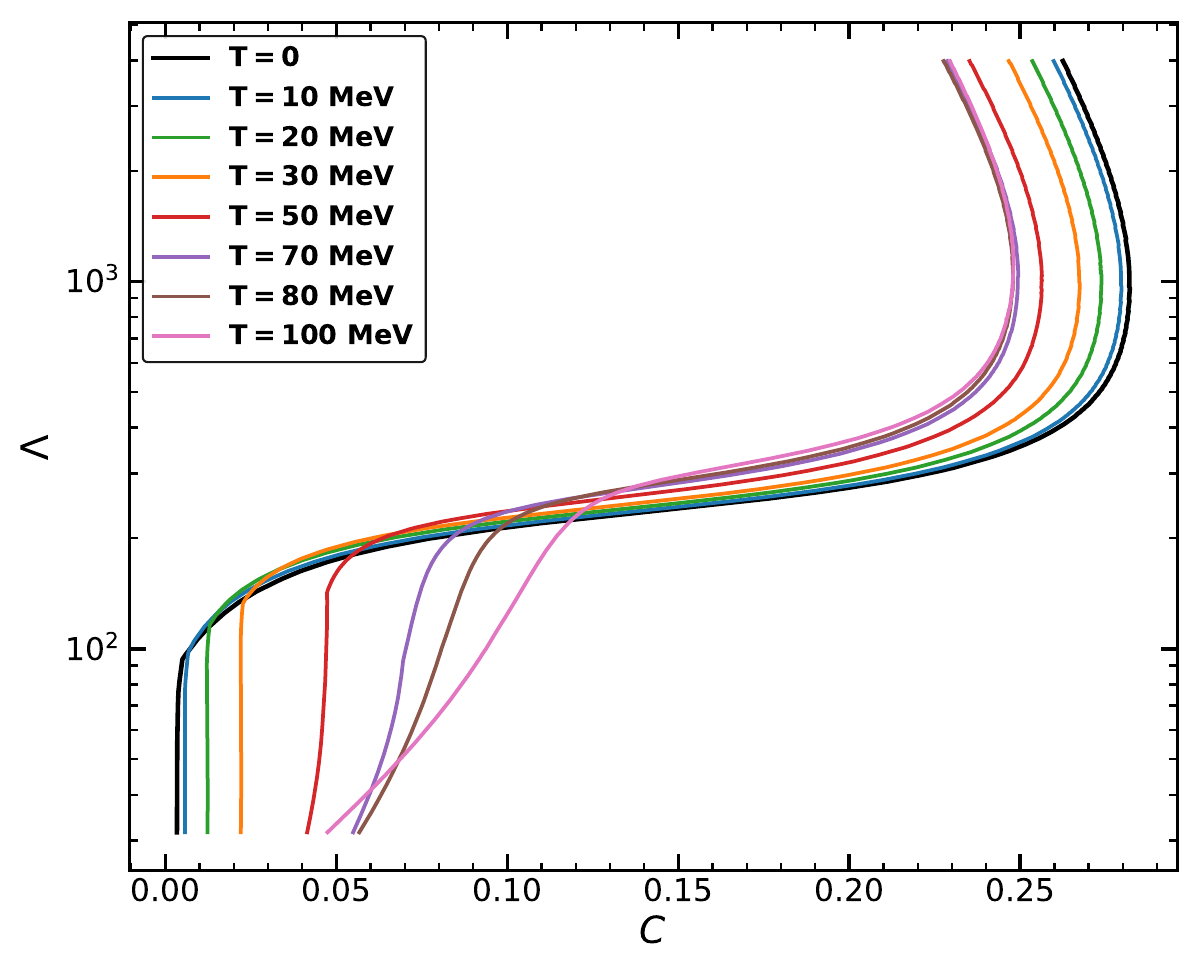}
\caption{The dimensionless deformability $\Lambda$ versus compactness parameter $C=\frac{M}{R}$ of axionic neutron star at different temperature T. The temperatures are taken as T=10, 20, 30, 50, 70, 80 and 100 MeV.}
% \textcolor{red}{Alt text: Plot of dimensionless tidal deformability versus stellar compactness for temperatures from 10 to 100 MeV, showing deformability decreasing as compactness increases, with $C_{max}$ rising toward about 0.28 before declining slowly.}}
\label{fig:lc}
\end{figure}
\\
%%%%%%%%%%%%%%%%%%%%%%%%%%%%%%%%%%%%%%%%%%%%%%%%%%
%%%%%%%%%%%%%%%%%%%%%%%%%%%%%%%%%%%%%%%%%%%%%%%%%%
%\subsection{The mass-radius relation:}
The compactness parameter $C_{max}=\frac{M_{max}}{R_{max}}$ of the hybrid star for different combinations of temperatures is shown in Fig.~\ref{fig:lc}. Also, the maximum $\Lambda_{max}$ and  $C_{max}$ as well as at the canonical values of $\Lambda_{1.4}$ and  $C_{1.4}$
are listed in Table II. From the table, it is evident that the canonical tidal deformability $\Lambda_{1.4}$ is maximum at T=30 MeV. The $R_{1.4}$ and $C_{1.4}$ are maximum and minimum at T=80 MeV. Similarly, the $R_{max}$ and $\Lambda_{max}$ T=80 MeV and T=30 MeV. The results of our calculations referred to the axionic star expanding and producing  the $\Lambda_{max}$ to a certain temperature and beyond which the expansion materialized. That observation reflects in  Fig.~\ref{fig:lc}. From Fig.~\ref{fig:lc}, one can see that the  $C_{max}$ increases gradually upto $C_{max}\sim 0.28$ and then decreases slowly. $\Lambda_{\rm max}$ behaves non-monotonically with temperature, peaking near $T = 30$ MeV before declining. This reflects a competition between two effects. Thermal expansion inflates $R$ and thus increases $\Lambda \propto R^5$. But the maximum-mass configuration softens at higher $T$, reducing $M_{\rm max}$ and the compactness. The canonical tidal deformability $\Lambda_{1.4}$ grows rapidly with temperature. It far exceeds the GW170817 bound $\tilde{\Lambda} \leq 800$ \cite{PRL119-161101-2017} for all finite-$T$ cases in Table~II. This confirms that the GW170817 constraint applies strictly to cold, $\beta$-equilibrated neutron stars. It does not apply to the hot neutron star configurations studied here.
\\
\\
%%%%%%%%%%%%%%%%%%%%%%%%%%%%%%%%%%%%%%%%%%%%%%%%%%
\section{Summary and Conclusions}
\label{sec:conclusion}
In summary, instead of a sharp hybrid equation-of-state as it is originally proposed \cite{PRL122-122701-2019}, the calculated results given in this paper are within a smooth and thermodynamically consistent quarkyonic-inspired crossover framework \cite{JCAP2025-056-2025,IJMPD35-2650015-2026}.  In this approach, the parameter $n_t$ is the central density of the crossover region associated with the onset of quarkyonic-inspired dynamics along with the trapped axion, which is a dark matter candidate. In the analysis, we have taken the crossover density $n_t = 0.3 ~\mathrm{fm}^{-3}$ and QCD confinement scale $\Lambda_{cs}=800$ MeV. 
In other words, for  $n_t = 0.3~\mathrm{fm}^{-3}$,
the  quark degrees of freedom are quite  important in the core of the NS and contribute considerably to the energy density and pressure support.
The nucleon medium follows the low and intermediate densities region, while the  quark degree of freedom enhance the pressure in the supranuclear densities beyond $n_t$. Since, the axions are trapped mostly in the quark medium due to supranuclear density along with an axionic distribution throughout the neutron star, we add the energy density and pressure contribution of axion to the resulting equation of state. The final EoSs are employed as input to the Tolman-Oppenheimer-Volkoff  equations, which yield different properties of neutron star configurations, such as $M_{max}$, $C_{max}$, $R_{max}$ and
$\Lambda_{max}$ along with the canonical values 
$C_{1.4}$, $R_{1.4}$ and $\Lambda_{1.4}$. The agreement of our predicted results with the present observational data are quite good. The mass of the NS at different temperatures are almost constant, but the maximum radius $R_{max}$ increases a lot with temperatures.  This results also support that quarkyonic-inspired EoS can support compact stars with masses exceeding $2.5\,M_{\odot}$.
\\
\\
The main outcome of the present study is the amalgam of three EoSs, i.e., quarkyonic EoS in the central region, baryonic EoS beyond the transition density $n_t$ till the surface in addition with the hypothetical axionic EoS throughout the supernova remnant. Although, we have taken the axionic EoS throughout the system, we find it's production is mainly in the middle part of the star ($5-11$ km). However, the effect of axion to the mass of the star at the central region of the system is significant. The mass of the star at finite temperature affected by $\sim 60\%$ in the center of the NS and it goes on decreasing with it's radius, because a lot of axions are absorbed near the vicinity of the center and the emission of axion flux is substantial on the crust and surface regions. This study is the first self-consistent calculation of thermal axion effects on neutron star structure within the quarkyonic-inspired smooth crossover framework. It shows that axion dark matter leaves a measurable fingerprint on the structural observables ($M$, $R$, $\Lambda$) and on the cooling history of the neutron star \cite{JCAP2025-056-2025,IJMPD35-2650015-2026}. The axion emissivity has an extreme $T^6$ sensitivity \cite{PRL53-1198-1984}. This means the axion cooling channel is most potent in the early hot phase of the neutron star's life. Next-generation neutrino detectors could resolve the time profile of the neutrino burst from a galactic supernova. This could constrain the axion-nucleon coupling $g_\theta$ more tightly than current SN~1987A bounds \cite{PRL60-1793-1988}. Extending this framework to include strange quark matter, hyperons, or a self-interacting axion potential would further enrich the physics. We leave this for future work.
%%%%%%%%%%%%%%%%%%%%%%%%%%%%%%%%%%%%%%%%%%%%%%%%%%
%%%%%
%%%%%%%%%%%%%%%%%%%%%%%%%%%%%%%%%%%%%%%%%%%%%%%%%%
\section{Acknowledgement}
% %%%%%%%%%%%%%%%%%%%%%%%%%%%%%%%%%%%%%%%%%%%%%%%%%%
The authors gratefully acknowledge Prof. D. Das for fruitful discussions and Prof. M. Bhuyan (Institute of Physics, Bhubaneswar) for a critical reading of the manuscript.

% %%%%%%%%%%%%%%%%%%%%%%%%%%%%%%%%%%%%%%%%%%%%%%%%%%
 
%\bibliographystyle{apsrev4-2}
%\bibliography{nstar}

\begin{thebibliography}{99}
\bibitem{PR45-138-1934} W. Baade and F. Zwicky, Phys. Rev. {\bf 45}, 138 (1934).
\bibitem{PR46-76-1934} W. Baade and F. Zwicky, Phys. Rev. {\bf 46}, 76 (1934).
\bibitem{ApJL848-L12-2017} B.~P.~Abbott {\it et al.} (LIGO Scientific, Virgo, and Partner Collaborations), Astrophys. J. Lett. {\bf 848}, L12 (2017).
\bibitem{PRL119-161101-2017} B.~P.~Abbott {\it et al.} (LIGO Scientific and Virgo Collaborations), Phys. Rev. Lett. {\bf 119}, 161101 (2017).
\bibitem{ApJL887-L24-2019} M.~C.~Miller {\it et al.}, Astrophys. J. Lett. {\bf 887}, L24 (2019).
\bibitem{ApJL918-L28-2021} M.~C.~Miller {\it et al.}, Astrophys. J. Lett. {\bf 918}, L28 (2021).
\bibitem{AA467-395-2007}R. Oechslin, H.-T. Janka, A. Marek, $A\&A$ {\bf 467(2)}, 395 (2007).
\bibitem{Nature403-727-2000} A. Burrows, Nature {\bf 403}, 727 (2000). 
\bibitem{NS-2003} B. Ryden. Neutron stars (2003). URL \url{http://www.astronomy.ohio-state.edu/~ryden/ast162_5/notes21.html}
\bibitem{PRL120-182701-2018}E.F. Brown, A. Cumming, F.J. Fattoyev, C.J. Horowitz, D. Page,
S. Reddy, Phys. Rev. Lett. {\bf 120}, 182701 (2018).
\bibitem{AA373-L17-2001}A.D. Kaminker, P. Haensel, D.G. Yakovlev, $A\&A$ {\bf 373(2)}, L17
(2001).
\bibitem{AA42-169-2004}D. Yakovlev, C. J. Pethick, Annu. Rev. $A\&A$ {\bf 42}, 169 (2004).
\bibitem{MNRAS363-555-2005}O. Y. Gnedin, M. Gusakov, A. Kaminker, D. G. Yakovlev,
Mon. Not. Roy. Astron. Soc. {\bf 363}, 555 (2005).
\bibitem{PRD98-103015-2018}K. Hamaguchi, N. Nagata, K. Yanagi, and J. Zheng, Phys. Rev. D {\bf 98}, 103015 (2018).
\bibitem{PRL40-279-1978}F. Wilczek, Phys. Rev. Lett. {\bf 40}, 279 (1978).
\bibitem{PRL40-223-1978}S. Weinberg, Phys. Rev. Lett. {\bf 40}, 223 (1978).\bibitem{PRD64-043002-2001} Naokia Iwamoto, Phys. Rev. D {\bf 64}, 043002 (2001).
\bibitem{PRL38-1440-1977} R. D. Peccei and H. R. Quinn, Phys. Rev. Lett. {\bf 38},
1440 (1977).
\bibitem{PRL43-103-1979}J. E. Kim, Phys.
Rev. Lett. {\bf 43}, 103 (1979).
%\bibitem{NPB166-493-1980}
%M. Shifman, A. Vainshtein, and V. Zakharov, Nucl. Phys.
%B166, 493 (1980).
\bibitem{PLB104-199-1981}M. Dine, W. Fischler, and
M. Srednicki, Phys. Lett. B {\bf 104}, 199 (1981).
\bibitem{ApJL934-L17-2022} R.~W.~Romani {\it et al.}, Astrophys. J. Lett. {\bf 934}, L17 (2022).
\bibitem{PRL60-1793-1988} M.~S.~Turner, Phys. Rev. Lett. {\bf 60}, 1793 (1988).
\bibitem{PRL53-1198-1984} Naoki Iwamoto, Phys. Rev. Lett. {\bf 53}, 1198 (1984).
\bibitem{PRC90-055203-2014} M. Dutra, O. Lourenco, S.S. Avancini, B.V.Carlson, A. Delfino, D.P. Menezes, C. Providencia, S. Typel, and J. R. Stone, Phys. Rev. C {\bf 90}, 055203 (2014).
\bibitem{PRC70-058801-2004} M. D. Estal, M.Centelles, X. Vi\~nas and  S.K. Patra, Phys. Rev. C {\bf 70}, 058801 (2004).
\bibitem{PRD9-3471-1974} A. Chodos, R. L. Jaffe, K. Johnson, C. B. Thorn, and V. F. Weisskopf, Phys. Rev. D {\bf 9}, 3471 (1974).
\bibitem{PRD10-2599-1974} A. Chodos, R. L. Jaffe, K. Johnson, and C. B. Thorn, Phys. Rev. D {\bf 10}, 2599 (1974).
\bibitem{PR122-345-1961} Y. Nambu and G. Jona-Lasinio, Phys. Rev. {\bf 122}, 345 (1961).
\bibitem{PR124-246-1961} Y. Nambu and G. Jona-Lasinio, Phys. Rev. {\bf 124}, 246 (1961).
\bibitem{PRD73-014019-2006} C. Ratti, M. A. Thaler, and W. Weise, Phys. Rev. D {\bf 73}, 014019 (2006).
\bibitem{AJ810-134-2015} T. Kl\"ahn and T. Fischer, The Astrophysical Journal {\bf 810}, 134 (2015).
\bibitem{PPNP33-477-1994}C. D. Roberts and A. G. Williams, Prog. Part. and Nucl. Phys. {\bf 33}, 477 (1994).
\bibitem{PRC91-035802-2015} M. Drews and W. Weise, Phys. Rev. C {\bf 91}, 035802 (2015).
\bibitem{PRL122-122701-2019} L. McLerran and S. Reddy, Phy. Rev. Lett. {\bf 122}, 122701 (2019).
\bibitem{JCAP2025-056-2025}D. Dey, J. A. Pattnaik, H. Das, A. Kumar, R. Panda, and S. Patra, Journal of Cosmology and Astroparticle
Phys. {\bf 2025 (01)}, 056 (2025).
\bibitem{PR55-364-1939} R. C. Tolman, Phys. Rev. {\bf 55}, 364 (1939).
\bibitem{PR55-518-1939} J. R. Oppenheimer, G. M. Volkoff, Phys. Rev. {\bf 55}, 518 (1939).
\bibitem{PS96-125319-2021}J. A. Pattnaik, M. Bhuyan, R. N. Panda, and S. K. Patra, Physica Scripta {\bf 96}, 125319 (2021).
\bibitem{CPC46-094103-2022} J. A. Pattnaik, R. N. Panda, M. Bhuyan, and S. K. Patra, Chinese Physics C {\bf 46}, 094103 (2022).
\bibitem{IJMPD35-2650015-2026} J. A. Pattnaik and S. K. Patra, Int. J. Mod. Phys. D {\bf 35}, 2650015 (2026).
\bibitem{PRD102-023021-2020} T. Zhao and J. M. Lattimer, Phys. Rev. D {\bf 102}, 023021 (2020).
\bibitem{PRC36-2590-1987} R. J. Furnstahl, C. E. Price, and G. E. Walker, Phys. Rev. C {\bf 36}, 2590 (1987).
\bibitem{NPA615-441-1997}R. Furnstahl, B. D. Serot, and H.-B. Tang, Nucl. Phys. A {\bf 615}, 441 (1997).

\bibitem{APJ170-299-1971} G. Baym, C. Pethick, and P. Sutherland, Astrophys. J. {\bf 170}, 299 (1971).
\bibitem{PRC55-540-1997} G. A. Lalazissis, J. K\"onig, and P. Ring, Phys. Rev. C {\bf 55}, 540 (1997).
\bibitem{JPG40-085104-2013} S. K. Singh, M. Bhuyan, P. K. Panda, and S. K. Patra, J. Phys. G: Nucl. and Part. Phys. {\bf 40}, 085104 (2013).
\bibitem{CPC45-025101-2021} O. Lourenc¸o, C. H. Lenzi, M. Dutra, T. Frederico, M. Bhuyan, R. Negreiros, C. V. Flores, G. Grams, and D. P.
Menezes, Chinese Physics C {\bf 45}, 025101 (2021).
\bibitem{PRL95-122501-2005}B. G. Todd-Rutel and J. Piekarewicz, Phys. Rev. Lett. {\bf 95}, 122501 (2005).
\bibitem{PRC97-045806-2018}B. Kumar, S. K. Patra, and B. K. Agrawal, Phys. Rev. C {\bf 97}, 045806 (2018).
\bibitem{NPA966-197-2017} B. Kumar, S. Singh, B. Agrawal, and S. Patra, Nucl. Phys. A {\bf 966}, 197 (2017).
\bibitem{ANP16-52-1986} B. D. Serot and J. D. Walecka, Adv. Nucl. Phys. {\bf 16}, 1 (1986).
\bibitem{PTEP2013-073D01-2013} K. Masuda, T. Hatsuda, and T. Takatsuka, Prog. of Theo. and Expt. Phys. {\bf 2013}, 073D01 (2013).
\bibitem{PRD100-103022-2019} S. Han, M. A. A. Mamun, S. Lalit, C. Constantinou, and M. Prakash, Phys. Rev. D {\bf 100}, 103022 (2019).
\bibitem{PRC74-055803-2006} T. K. Jha, P. K. Raina, P. K. Panda, and S. K. Patra, Phys. Rev. C {\bf 74}, 055803 (2006).
\bibitem{PRC75-035808-2007} B. K. Sharma, P. K. Panda, and S. K. Patra, Phys. Rev. C {\bf 75}, 035808 (2007).
\bibitem{BTFT925-1-2016}M. Laine and A. Vuorinen, “Basics of Thermal Field Theory,” Lect. Notes Phys. {\bf 925}, 1 (2016).
%\bibitem{UNIVERSE11-2025}A. Salvio, “Introduction to Thermal Field Theory: From First Principles to Applications,” Universe
%11, 16 (2025) [arXiv:2411.02498 [hep-ph]].
\bibitem{arxiv:2506.1692v2}Momchil Naydenova
and Alberto Salvio,  arXiv:2506.16932v2 [hep-ph] 18 Aug 2025.
\bibitem{NPB166-493-1980}
M. Shifman, A. Vainshtein, and V. Zakharov, Nucl. Phys.
B {\bf 166}, 493 (1980).
\bibitem{YF31-497-1980}A. P. Zhitnitskii, Yad. Fiz. {\bf 31}, 497 (1980) [Sov. J. Nucl. Phys.
{\bf 31}, 260 (1980)].
\bibitem{PRL128-091102-2022}M. Buschmann,  C Dessert, J. W. Foster, A. J. Long and B. R. Safdi, Phys. Rev. Lett. {\bf 128}, 091102 (2022).
%%%%%%%%%%%%%%%%%%%%%%%%%%%%%%%%%%%%%%%%%%%%%%%%%%%%%%%%%%%%%%%%%%%%%%%%%%%%%%%%%%%%%%%%%%%%%%%%%%%%%%%%%%%%%%5
%%%%%%%%%%%%%%%%%%%%%%%%%%%%%%%%%%%%%%%%%%%%%%%%%%%%%%%
%%%%%%%%%%%%%%%%%%%%%%%%%%%%%%%%%%%%%%%%%%%%%%%%%%%5%
\bibitem{reed21}
B. T. Reed, F. J. Fattoyev, C. J. Horowitz, and J. Piekarewicz,
Phys. Rev. Lett. \textbf{126}, 172503 (2021).

\bibitem{Bethe_1971}
H. A. Bethe,
Ann. Rev. Nucl. Sci. \textbf{21}, 93 (1971).
\bibitem{shlomo06}
S. Shlomo, V. M. Kolomietz, and G. Col\`o,
Eur. Phys. J. A \textbf{30}, 23 (2006).
\bibitem{colo08}
G. Col\`o,
Phys. Part. Nucl. \textbf{39}, 286 (2008).
\bibitem{garg18}
U. Garg and G. Col\`o,
Prog. Part. Nucl. Phys. \textbf{101}, 55 (2018).
\bibitem{Colo_2014}
G. Col\`o, U. Garg, and H. Sagawa,
Eur. Phys. J. A \textbf{50}, 26 (2014).
\bibitem{Zimmerman_2020}
J. Zimmerman, Z. Carson, K. Schumacher, A. W. Steiner, and K. Yagi,
arXiv:2002.03210 [astro-ph.HE] (2020).
\bibitem{Stone_2014}
J. R. Stone, N. J. Stone, and S. A. Moszkowski,
Phys. Rev. C \textbf{89}, 044316 (2014).
\bibitem{Pearson_2010}
J. M. Pearson, N. Chamel, and S. Goriely,
Phys. Rev. C \textbf{82}, 037301 (2010).
\bibitem{TLi_2010}
T. Li \textit{et al.},
Phys. Rev. C \textbf{81}, 034309 (2010).

% \bibitem{TEU69-1990}E. W. Kolb and M. S. Turner, “The Early Universe", Frontiers in Physics, Vol. 69, Addison-Wesley,
% Redwood City, CA, 1990.
% \bibitem{LAP2019}S. Weinberg, Lectures on Astrophysics, Cambridge University Press; 2019.
\end{thebibliography}

\end{document}